%% file: main.tex
\documentclass[sigconf,screen]{acmart}

\AtBeginDocument{%
  \providecommand\BibTeX{{%
    \normalfont B\kern-0.5em{\scshape i\kern-0.25em b}\kern-0.8em\TeX}}}

\setcopyright{acmlicensed}
\acmPrice{15.00}
\acmDOI{10.1145/3468264.3468599}
\acmYear{2021}
\copyrightyear{2021}
\acmSubmissionID{fse21main-p575-p}
\acmISBN{978-1-4503-8562-6/21/08}
\acmConference[ESEC/FSE '21]{Proceedings of the 29th ACM Joint European Software Engineering Conference and Symposium on the Foundations of Software Engineering}{August 23--28, 2021}{Athens, Greece}
\acmBooktitle{Proceedings of the 29th ACM Joint European Software Engineering Conference and Symposium on the Foundations of Software Engineering (ESEC/FSE '21), August 23--28, 2021, Athens, Greece}

\usepackage{lookup}
\input{lookup_table}

\input{utils.tex}

\input{aliases.tex}

\begin{document}

\title[Reel Life vs. Real Life: How Software Developers Share Their Daily Life through Vlogs]{Reel Life vs. Real Life: \\ How Software Developers Share Their Daily Life through Vlogs}

\author{Souti Chattopadhyay}
\affiliation{%
  \institution{Oregon State University}
  \city{Corvallis}
  \state{OR}
  \country{USA}
}
\email{chattops@oregonstate.edu}

\author{Thomas Zimmermann}
\affiliation{%
  \institution{Microsoft Research}
  \city{Redmond}
  \state{WA}
  \country{USA}
}
\email{tzimmer@microsoft.com}

\author{Denae Ford}
\affiliation{%
  \institution{Microsoft Research}
  \city{Redmond}
  \state{WA}
  \country{USA}
}
\email{denae@microsoft.com}

\renewcommand{\shortauthors}{Souti Chattopadhyay, Thomas Zimmermann, Denae Ford}

\begin{abstract}

Software developers are turning to vlogs (video blogs) to share what a day is like to walk in their shoes. Through these vlogs developers share a rich perspective of their technical work as well their personal lives. However, does the type of activities portrayed in vlogs differ from activities developers in the industry perform? Would developers at a software company prefer to show activities to different extents if they were asked to share about their day through vlogs? To answer these questions, we analyzed 130 vlogs by software developers on YouTube and conducted a survey with 335 software developers at a large software company. We found that although vlogs present traditional development activities such as coding and code peripheral activities (11\%), they also prominently feature wellness and lifestyle related activities (47.3\%) that have not been reflected in previous software engineering literature. We also found that developers at the software company were inclined to share more non-coding tasks (e.g., personal projects, time spent with family and friends, and health) when asked to create a mock-up vlog to promote diversity. These findings demonstrate a shift in our understanding of how software developers are spending their time and find valuable to share publicly. We discuss how vlogs provide a more complete perspective of software development work and serve as a valuable source of data for empirical research.
\end{abstract}



\maketitle

\section{Introduction}

Software development is a lucrative and popular career option, and incites curiosity even in the general public unfamiliar with programming. Some developers have created vlogs (video blogs) to showcase their entire day including coding sessions, team meetings and even life after work-hours. These vlogs present challenges they face when working on a project as well as the victories of new achievements on personal side projects. Likewise, outside of technical work, vlogs by software developers provide a personal glimpse into various careers in software development industry.
The transparency that these vlogs provide a broader perspective of what it means to be a developer---one helpful in inspiring the next generation of developers.

Although these vlogs are helpful in painting a picture for a wider general audience, how are they similar to what developers in industry would want as their online image?
Do developers in industry engage in activities that are different from how it is currently portrayed in vlogs?
In this work, we use vlogs to understand what activities developers find valuable and want others to know about their professional and personal life. Vlogs provide a richer medium than text-based blogs where creators are limited by the format of content they can share. Vlogs allow creators tell a story of their life via video and audio that reflect their personalities~\cite{aran2013broadcasting}. This creates gives the audience the feel of physically shadowing the developer without having to do so.
Additionally, the succinct nature of vlogs encourages vloggers to be selective in what content they will be sharing---often times optimizing for what they think will be the most valuable content to show publicly. Understanding the online image of developers can help gain a better understanding of what software developers find valuable to share as part of their image, and how the general public perceives software developers. %

A popular format of developer vlogs are ``A day in the life of a developer'' videos. These vlogs mirror a developer's entire day---from waking up, the food they eat, going to work, what they do at work (like coding, testing, meetings, co-working in teams), breaks, and what they do outside of work (working out, going to movies, spending time with kids and families, or playing games). Each vlog has its own story and emphasize some of these activities.\footnote{To get an idea of the content of these vlogs, we recommend the reader to watch the following examples: \begin{enumerate} \item ``Day in the Life of a Software Engineer (First week!)'' (505K views) \newline \url{https://youtu.be/bX8hvldRx1M}; \item ``Day in the Life of a Software Engineer | New York'' (281K views) \newline \url{https://youtu.be/qZx7pvRootk}; and \item ``Day of Amazon Software Developer'' (519K views) \newline \url{https://youtu.be/c8dd9f5MamU} \end{enumerate}}

To identify how developers would present their day, we conducted a mixed-methods study with two populations of software developers: 1) those who vlog and 2) those from a large software company (Section~\ref{sec:methodology}). First, we analyzed vlogs from 130 ``\emph{day in the life of a software developer}'' videos on \yt. Then we qualitatively analyzed these activities in these videos and the percentage of time spent in each activity. To understand how this distribution of time spent in videos differs between the two populations of developers, we designed and conducted survey at a \bigcompany. We received responses from 335 developers. In this survey we included questions about the time spent in specific activities, frequency of what they would show in a pretend vlog, and activity usage on social media platforms. %

The results of our study demonstrate a fuller perspective of the life of a software developer (Section~\ref{sec:results}). 
We find that in addition to coding and peripheral coding tasks like debugging, \vlogdevs spend a significant amount of time spent on non-work related elements of life such as lifestyle, spending time with loved ones, and dedicating time to health.
In contrast, large software company developers (\MSdevs) overall reported that they would present less time showcasing these activities and reported that would dedicate more time problem solving tasks and coding-related work. We also find that the percentage  of how much of the non-code related activities varied based on the motivation for creating the \pretendvlog. For example, \MSdevs who were presented with the motivation of supporting diversity found it valuable to show more content on personal projects, interactions with family and friends, personal brand, and health.
From these findings, we discuss (Section~\ref{sec:discussion}) how we can operationalize this new understanding of how developers spend their time and provide implications for how vlogs can be used as a dataset for understanding a range of developer perspectives, and various insights that developers, managers, and the SE community can gain from them.

\section{Research Questions}

To understand a broader perspective of how software developers, work we designed a study guided by three research questions: 
\begin{description}
\setlength{\itemsep}{0pt}
\setlength{\parskip}{2pt}
    \item[\textbf{RQ1}] \textbf{\RQA} Similar to a study of shadowing developers, vlogs by developers on \yt provide a rich source of data on how developers present their daily routines. But vlogs allow developers to control their own narrative and online avatar, thus giving a peek into their ideal life.
    Through this research question, we wanted to understand how developers portray their image through the different activities shown in their vlogs that characterizes their day and, in essence, themselves. Through qualitative analysis of 130 vlogs from developers, we categorize daily activities and respective screen-time representing the activities.

    \item[\textbf{RQ2}] \textbf{\RQB} Next, we wanted to understand how much time developers spend in reality across these activities through a survey of 335 developers in a large company. We further investigate the differences in the real time spent and the screen-time dedicated for activities and it's implications on software developer companies.

    \item[\textbf{RQ3}] \textbf{\RQC} Finally, to learn about the differences and diversity in what developers find valuable to portray their identify, we ask the developers from the company about how they would make vlogs. Through this analysis, we identify the impact of diversity and motivation on what aspect developers find valuable.
\end{description}

\section{Methodology}
\label{sec:methodology}

To answer the research questions, we conducted a mixed-methods study of analyzing data of software developers online (Section~\ref{sec:methodology:vlogs}) and data from a survey among developers who work at a large software development company (Section~\ref{sec:methodology:survey}). All study materials can be found in supplemental materials~\cite{supplementarymaterials}.

\subsection{Gathering Public Developer Presentation Through Vlogs}
\label{sec:methodology:vlogs}

To gather, how developers portray themselves publicly, we collected and analyzed vlog (video blog) data from YouTube, which is a popular online video-sharing platform. 
We focused on videos that are specifically about the day in the life of software engineers.
For the analysis, we identified 130 vlogs with a total length of 1,070 minutes and over 35 million views.

\subsubsection{Data Collection} We searched \yt to identify an initial set of 163 vlogs using a stratified sampling method to represent countries/regions with a strong developer presence. We identified the initial set of videos through \yt search using a combination of keywords with \emph{`developer life', `day in life', `day in life + software engineer'}. Resulting videos represented developers from North \& South America and Europe. To extend the representation to Asia, we conducted a focused search by adding South Korea, Japan, and India to our keywords. The resulting search also captured developer vlogs from other Asian countries like Singapore, Bangladesh, and China. We further identified 8 more videos through snowballing from (1) \yt recommendations that revealed vlogs that were not part of the initial sample, and (2) other vloggers mentioned in our sample of vlogs.

Our final sample consisted of 130 videos after excluding 33 videos from initial analysis (6 videos not in English, 14 videos captured more than once, 7 videos not by real developers as per the description in their channel, and 6 videos were different kinds like a recorded  conference talk. Throughout the rest of the papers we reference these vlogs as V1-V130.

\subsubsection{Data Preprocessing} To analyze the breadth of the videos, we manually identified and compiled standard \yt metadata from each videos including the date of upload, length of the video, location, subscriber count, view count, likes and dislikes count, and the number of comments. 
All 130 videos were uploaded within the last 4 years (January, 2016 - May, 2020); and are from 113 distinct developers. 103 videos were uploaded by men, 24 videos by women (1 featuring a couple, and 2 other videos we couldn't determine the gender). We should note the genders reported are in fact perceptible gender and are only determined to report the breadth of vloggers in our sample.
These vlogs come from 21 countries across the four continents of North and South America, Europe, and Asia. The countries with the most videos were USA, UK, Canada, India, and South Korea. The average length of the videos is 8.23 minutes, which is comparable to the typical length of \yt videos~\cite{clement2019youtube}.

To determine the reach of the vlogs, we analyzed the metadata associated with viewer engagement. Information within these vlogs reach a large audience. On average, each vlog had 270,800 views (min. = 83, max. = 7.131 million). Combined, these vlogs had 35 million views. 
It is difficult to report demographics of the viewers since \yt does not publicly provide any statistics about the viewers.

\subsubsection{Data Analysis}
\label{sec:vlog-analysis}

To answer how software vloggers present their daily routine in vlogs (\textbf{RQ1}), the first author transcribed the videos and assigned descriptive codes (labels/short phrases)~\cite{Saldana2009} to the various topics covered by vloggers as well as the activities they show as part of their everyday life. In the transcripts, for each \emph{activity} the start and end times in the video were recorded as well. We refer to the time an activity is shown in a video as \emph{vlog time}. Note that the vlog time is different from the actual time a developer spends on activity throughout their day (see the discussion in Section~\ref{sec:discussion:picture-perfect}). 
The first author recorded a total of 1135 activity codes across the videos.
All three authors then collaboratively reorganized these codes and performed selective coding~\cite{Saldana2009} by grouping related topics into stand-alone thematic concepts. The authors met multiple times over the next few weeks merging, splitting, and reorganizing the topics to identify the themes that describe ``what is a developer's life about'' as expressed in the vlogs. 

To analyse the different motivation(s) behind each vlog, the authors identified three motivations based on the prominent message of the vlog. 
First, developers in 29 videos were motivated to build and expand their own personal network and ventures  which included promoting themselves or their companies\citeVideo{M15}, the products they built, and other projects they engaged in like podcasts, talk shows etc. Second, in 30 vlogs, developers wanted to promote gender, race and ethnic diversity in software engineering and other computing fields by encouraging and associating with movements and events like \#womenintech, ``chick-tech'', ``BlackGirlsCode'' \citeVideo{M100}. And finally, in 94 vlogs, developers were motivated to create awareness about software engineering as a career choice by discussing how to pursue careers in computing, what expectations should people have about lifestyle and work, and the comforts and adjustments of pursuing a career in computing \citeVideo{M20}.

\subsection{Software Company Developer Survey}
\label{sec:methodology:survey}

To gather a perspective from a traditional experience of software developers, we designed and distributed a survey within a large software company.
We conducted a survey with 335 developers at a large software company (LSC) to understand the common activities they engage in throughout the day. Developers who post vlogs are more diverse in terms of job roles and types compared to the population of developers that researchers typically study. Through the survey, we compare our understanding of how developers spend time on different activities at work and beyond with our current knowledge of the various developer activity studies.
We will refer to the developers at this large software company from our survey as \emph{\MSdevs} throughout this paper.

\subsubsection{Survey Design} The survey focused on the online presence of software developers. Several questions were centered around a list of 14 activities that developers do during a day in their life. The list of 14 activities was curated from the ten categories of activities that were identified in the vlogs (discussed in Section~\ref{sec:result-rq1}). From the ten categories, we enumerated 14 different activities in the survey by splitting big categories such as ``Health, Workout, Lifestyle'' into separate activities, and we combined a few activities to better align and facilitate comparisons with activities that have also been identified in research related to developer workdays. 

The survey consisted of three parts. 
The first part consisted of demographic questions, including questions about general social media usage.
The second part asked developers how much time do they spend doing each activity, and then how much time they think ``other developers'' spend doing each of those activities \footnote{The analysis comparing the how much time developer think they spend on an activity and how much they think colleagues spend is included in the supplemental material~\cite{supplementarymaterials}.}
Finally, the third part asked developers that if they were to shoot a video of their daily life for a vlog, which of activities would they show the same, less, or more than their usual day. 

\smallskip
Question: \emph{``Typically, in any day of your life, how much time do you spend doing these activities?''}
We asked developers to estimate the the \textbf{time spent} in the 14 different activities on a ``typical day''.  For each of these activities, developers could choose one of the following time intervals: None, $< 15$ mins, $15-29$ mins, $30$ mins $- 1$ hour, between $1-2$ hours, $2-4$ hours, and $>4$ hours.

\smallskip
Question: \emph{``You and your team are working on a video about "A day in the life of a software engineer". Through the video, you want to (build and expand your personal brand | create awareness about software engineering as a career choice | promote diversity in software engineering). Compared to how frequently these activities occur in your life, how would you show these activities in the video?''}

This question about a ``\textbf{\pretendvlog}'' draws from study design techniques in participatory design research~\cite{elizarova2017participatory}. We included the question to collect bespoke, anticipatory experiences of what developers would be interested in sharing in their vlog. Asking about the content they would show and the relative duration provides us with an understanding of what developers find valuable and feel comfortable showcasing about the developer life publicly.

Developers were randomly assigned one of the three motivations to shoot the video: personal brand (112 developers), creating awareness (117) and promoting diversity (106). These three motivations were identified from the analysis of the 130 YouTube videos as the main reasons behind why developers post their ``day in the life'' vlogs.

\subsubsection{Survey Distribution} The developers were identified from a multinational company's internal address book contacts through their job roles. The survey was distributed through a personalized email invitation which included a link to the anonymous survey. After completion of the survey, respondents could enter a sweepstakes to win one of four USD \$100 gift cards. The survey was sent to 1,500 people, of which 335 people responded (response rate of 22.3\%, comparable to other surveys in software engineering~\cite{smith13}). 

The demographics of the respondents were as follows.
\emph{Age.} 15\% were 18-24 years, 28\% were 25-29 years, 20\% were 30-34 years, and 28\% were 35 - 44 years, and 9\% were 45 years or older.
\emph{Gender.} 6\% identified as a woman, 79\% as a man, and 1\% as non-binary / gender diverse, and 4\% preferred not to answer.
\emph{Social media usage.} Of the respondents, 68\% had previously watched YouTube videos related to ``developers'' or ``software development''. The social media platforms used the most were LinkedIn (81\% of respondents), YouTube (79\%), Facebook (61\%), Instagram (47\%), and Twitter (42\%). Developers used the social media platforms mainly to browse through others' posts (74\%), stay updated about colleagues and people they know (66\%), and to network with others (51\%). They also used social media to share technical skills and career advice (24\%) and build personal brands (14\%).

\subsubsection{Data Analysis} 
\label{sec:survey-analysis}
To answer \textbf{RQ2} and \textbf{RQ3}, we report and compare descriptive statistics in Sections~\ref{sec:result-rq2} and~\ref{sec:result-rq3}. 
To compare how different motivations change the activities that developers would show more prominently in vlogs, we performed Fisher Exact Value tests for each activity and motivation.

\subsubsection{Key Terminology}
To answer our research questions, we compare software developer experiences from vlogs and at a large software company. Below we provide the descriptions for key terms used for the comparisons in this paper:
\begin{itemize}
    \item \emph{Vlog} refers to the 130 YouTube videos (RQ1). 
    \item \emph{Vlog time} refers to the time an activity is shown in a video in the analysis of the YouTube videos (RQ1). 
    \item \emph{Reported time} refers to the time \MSdevs spend on an activity based on self-reported data from the survey (RQ2).
    \item \emph{\pretendvlog} refers to the videos that \MSdevs ``created'' as part of the participatory design question in the survey (RQ3).
\end{itemize}

\section{Results}
\label{sec:results}

We answer three research questions by expanding data analysis described in Sections~\ref{sec:vlog-analysis} and \ref{sec:survey-analysis}.

\subsection{\RQA (RQ1)}
\label{sec:result-rq1}

To better understand how developers present themselves online, we analyze existing self-presentations in the form of vlogs. From our qualitative analysis process outlined in Section~\ref{sec:vlog-analysis}, we found 10 prominent activity categories as presented in Table~\ref{tab:vlog-activities}. Next we describe the 10 categories. %

\input{tables/vlog_activities}

\newcommand{\activitystats}[2]{(#1 videos, #2\% of total time)}
\newcommand{\activitysection}[3]{\smallskip\textbf{#1 (#2 videos, #3\% vlog time):}\xspace}
\newcommand{\activitysectionnew}[4]{\smallskip\myiconinline{#4}\noindent\textbf{#1 (#2 videos, #3\% vlog time):}\xspace}

\activitysectionnew{Code sessions}{63}{11.1}{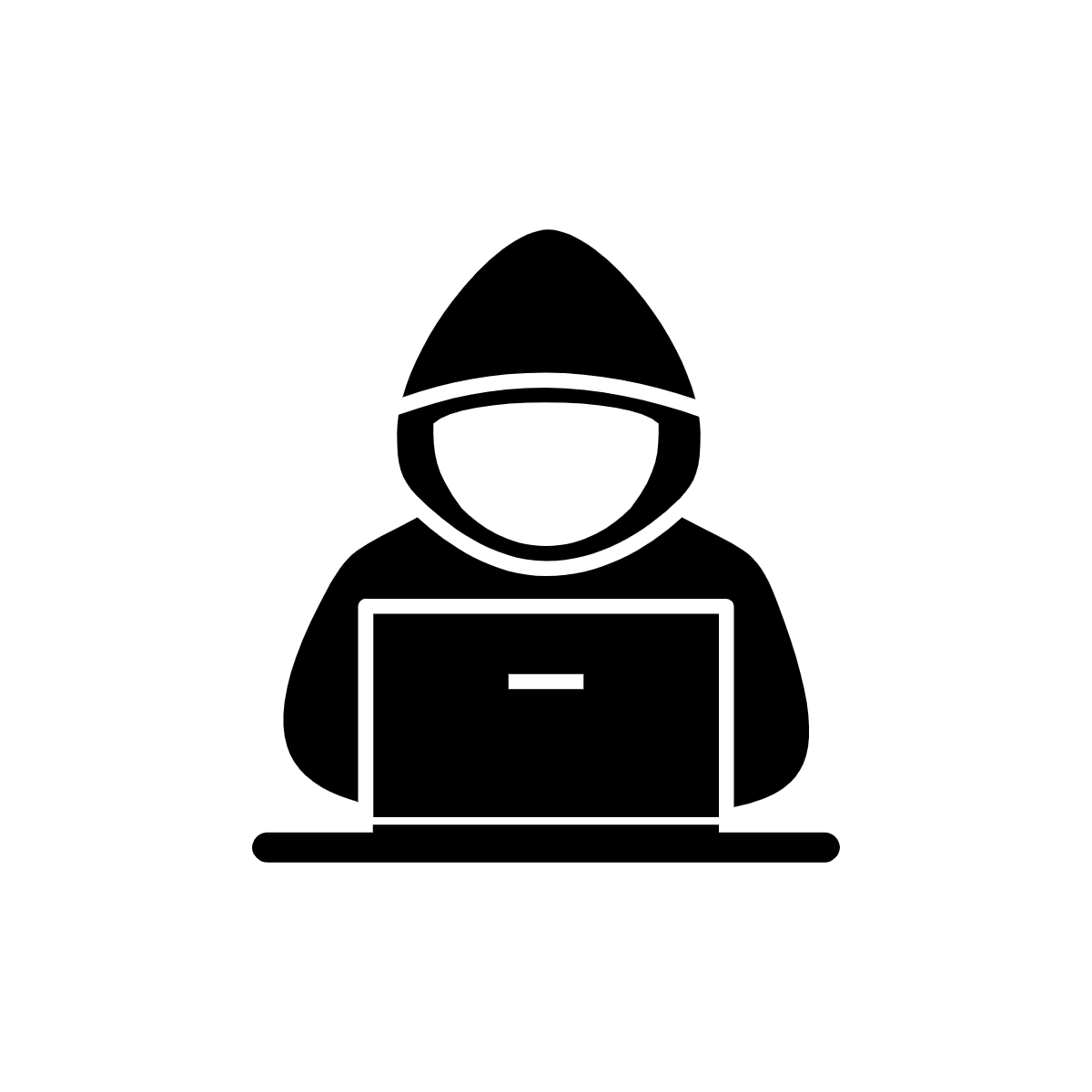} 
As anticipated, developers present themselves doing code and peripheral coding tasks as part of their work in the vlogs. There are various strategies to show coding, mostly this features as a timelapse of the coding session with their screens blurred. Once developers identify the task they will work on [described in managing work], they need to familiarize themselves with the relevant artifacts and code. They either learn the concepts upfront like ``authentication''\citeVideo{M1_42}, or they learn as they work on the task which includes implementing, testing [\lookupVideo{M1_15}, \lookupVideo{M1_92}], debugging \citeVideo{M1_29}, running and building as well as releasing and deploying code to provide new features or fix bugs [\lookupVideo{M1_38}, \lookupVideo{MS_16}]. They also show activities such as reading concepts related to a task [\lookupVideo{M1_03}, \lookupVideo{MS_2}], designing and sketching out the solution \citeVideo{M1_64}, searching solutions on Stack Overflow or GitHub [\lookupVideo{M1_19}, \lookupVideo{M2_100}], and working with repositories \citeVideo{M1_54}. 

When not directly implementing, developers review code which \inlinequote{basically is a commit that somebody submitted and they are sent for review in our team just to
figure out if there are any mistakes and errors that happen in the code.} \citeVideo{M1_08}. Code reviews feature frequently in the videos. Sometimes, code reviews are high priority tasks \citeVideo{M1_29}.  Videos also showed reviewing pull requests from other teams \citeVideo{M1_29}, interns \citeVideo{M1_20}, and junior members of the team. As a senior software engineer put it, \inlinequote{this is how I spends 60\% of my time} \citeVideo{M2_130}.

\activitysectionnew{Unspecified work}{54}{8.6}{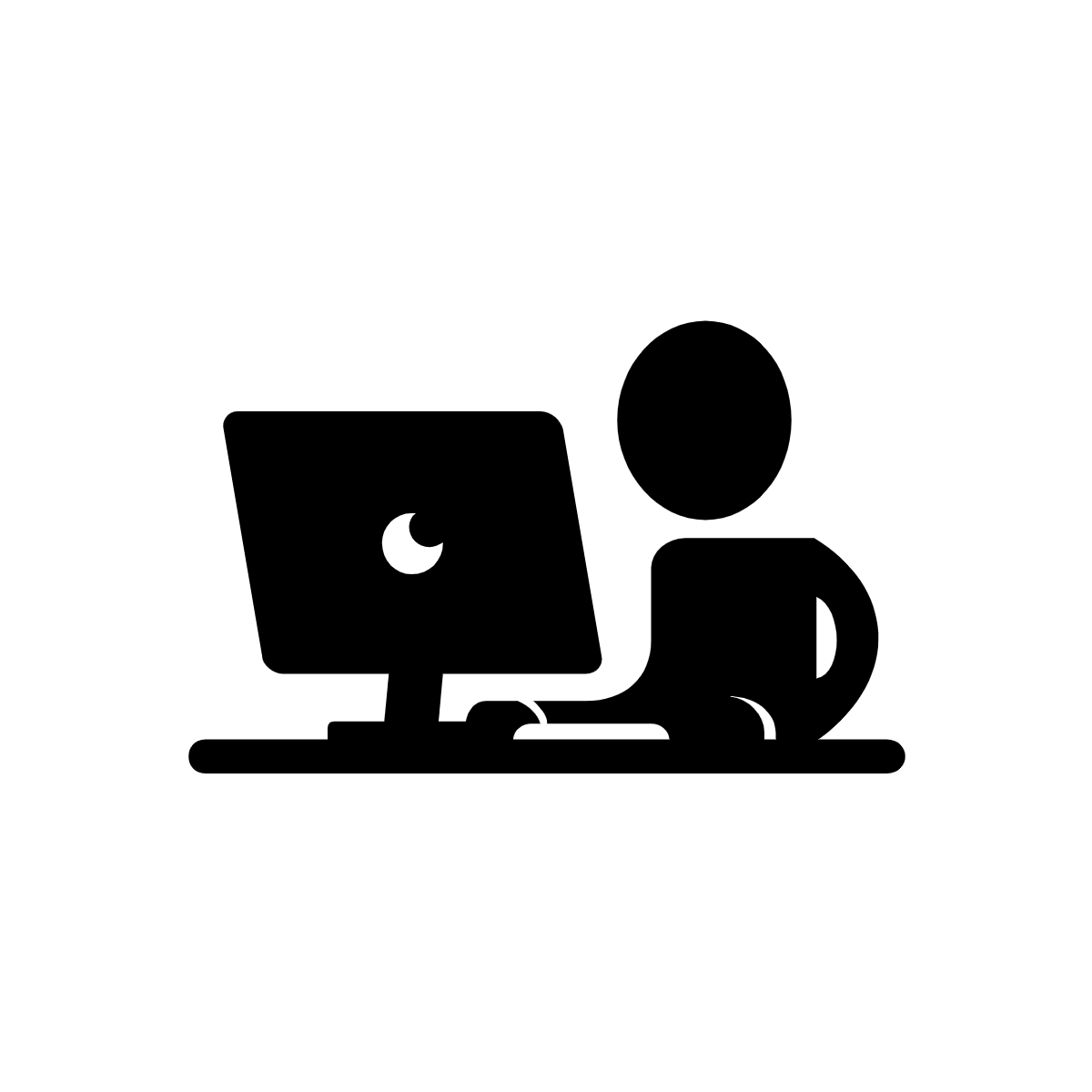} 
Developers in 54 videos did not specify what work they did within their office due to company policies or just privacy preference. During these times, the camera was usually left on and it captured a timelapse footage of the developer at their desk/office, typically showing the developers' face or scenery outside the window.

\activitysectionnew{Administrative Tasks}{19}{2.0}{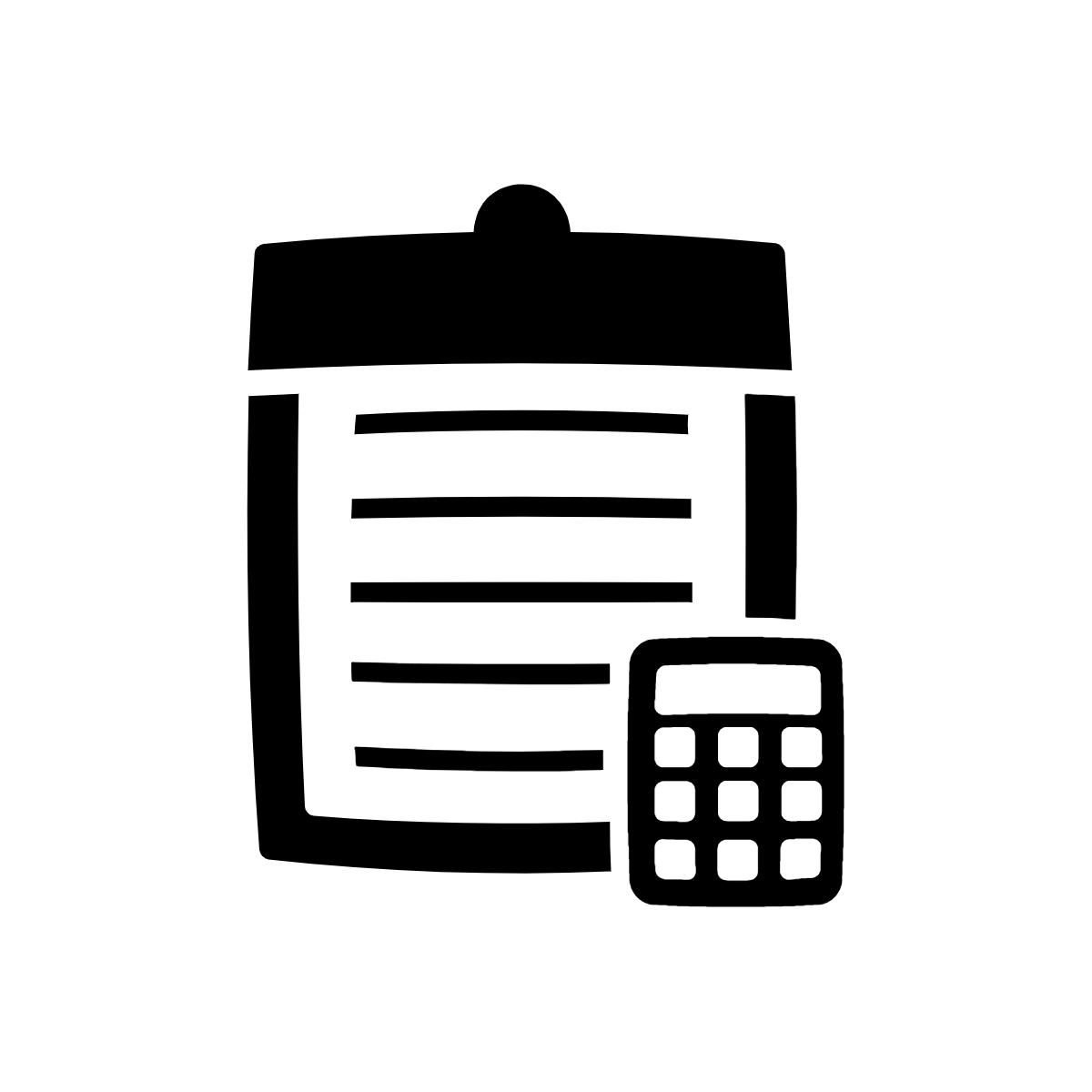} 
Contrary to belief, coding makes only a part of the activities. A developer reports \inlinequote{in terms of time I spend actually in the editor, I don't think it's that much. Maybe a couple hours per day} \citeVideo{M2_99}. Vlogs show developers spend time interviewing potential employees. For example, a tech lead of the company was \inlinequote{interviewing [today] for an on-site interview for general tech. I'm not film this. That's obviously not appropriate} \citeVideo{M1_05}. After the interviews, he goes on to give an update that he had \inlinequote{debriefs for two different interview candidates, discuss feedback with the other interviewers so that took up an hour of my afternoon.} \citeVideo{M4_144} shows how developers need to prepare for interviews they have to conduct, in his case it was the next day. 

Developers also meet often with interns and discuss their assignment and progress [\lookupVideo{M1_14}, \lookupVideo{M1_60}]. Administrative tasks become part of developer job, \emph{as you grow you also start to manage some people and mentor some people and look after small to mid-size teams. Every engineer inevitably is gonna do a little bit of management} \citeVideo{M2_123}. Once developers switch over to management roles, they spend time creating and assigning tasks, meeting with designers and architects to finalize designs, writing specification requirements for products \citeVideo{M2_133}, and other documentation like manuals outlining administrative tasks that describe policies and procedures that affect day-to-day operations of the team \citeVideo{M3_137}. 

Vlog also show administrative tasks like freelance developers need to manage their own clients and projects, and do additional administrative work like recording and reporting their own time \citeVideo{M1_54}, sending out invoices for their work \citeVideo{M1_09}, and reporting status of their work \citeVideo{M1_54}.

\activitysectionnew{Managing work/Planning the day}{34}{3.2}{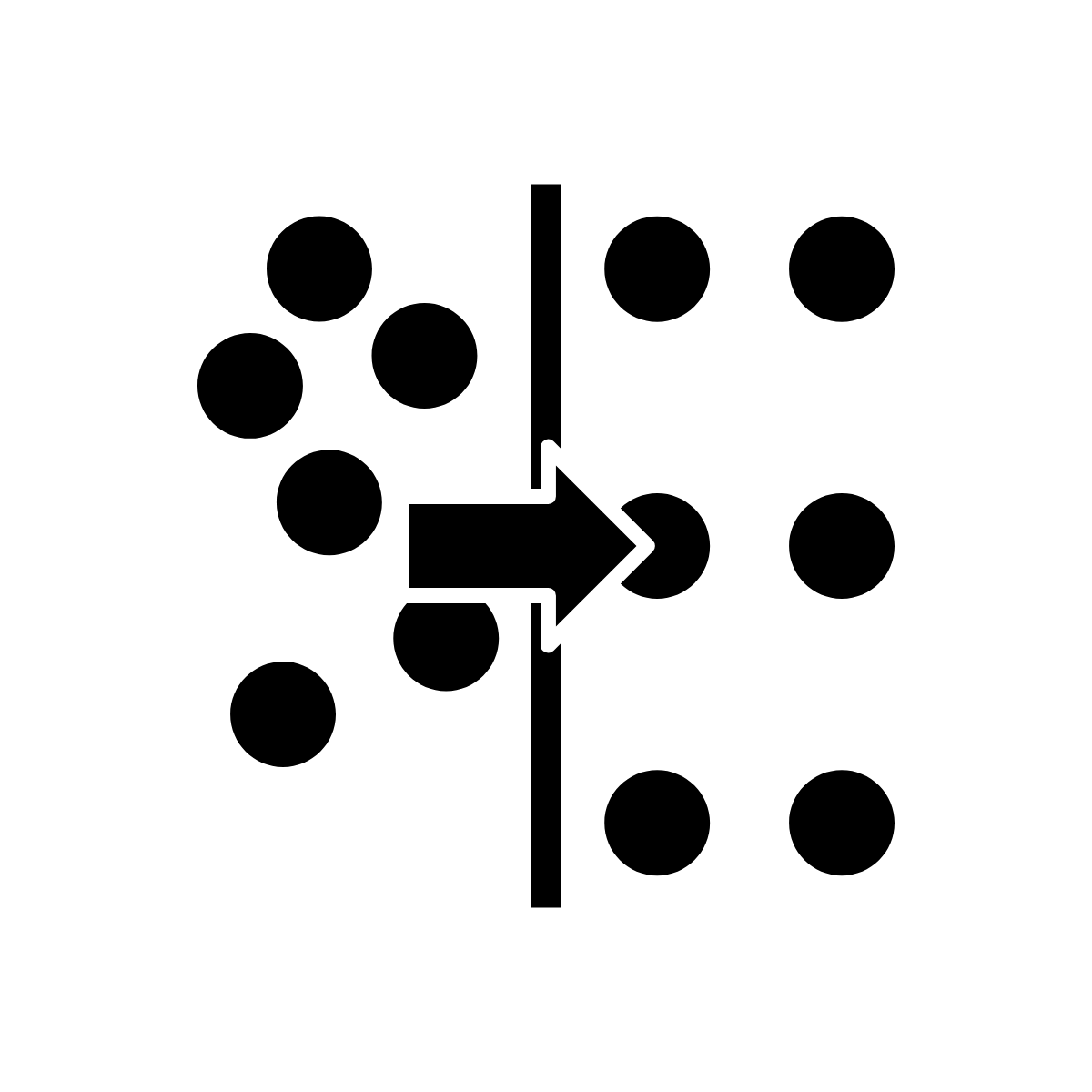} These snippets contain developers organizing their workday on paper journals \citeVideo{M2_125} or application like OmniFocus \citeVideo{M1_33} to make plan the day and make ToDo list, scheduling time for certain activities and estimating time \citeVideo{M1_85}. Developers who worked freelance used this time to manage clients, and developers at a company dealt with JIRA~\cite{JIRA} or other project management software. Usually developers start their day with these activities. However, few of them organize their next day before going to sleep \citeVideo{M1_02}, while some use this time to reflect on their previous day \citeVideo{M1_42}. Developers show different techniques and applications (like Notion \cite{notion}) they use to plan a productive day.

\activitysectionnew{Meetings}{65}{8.4}{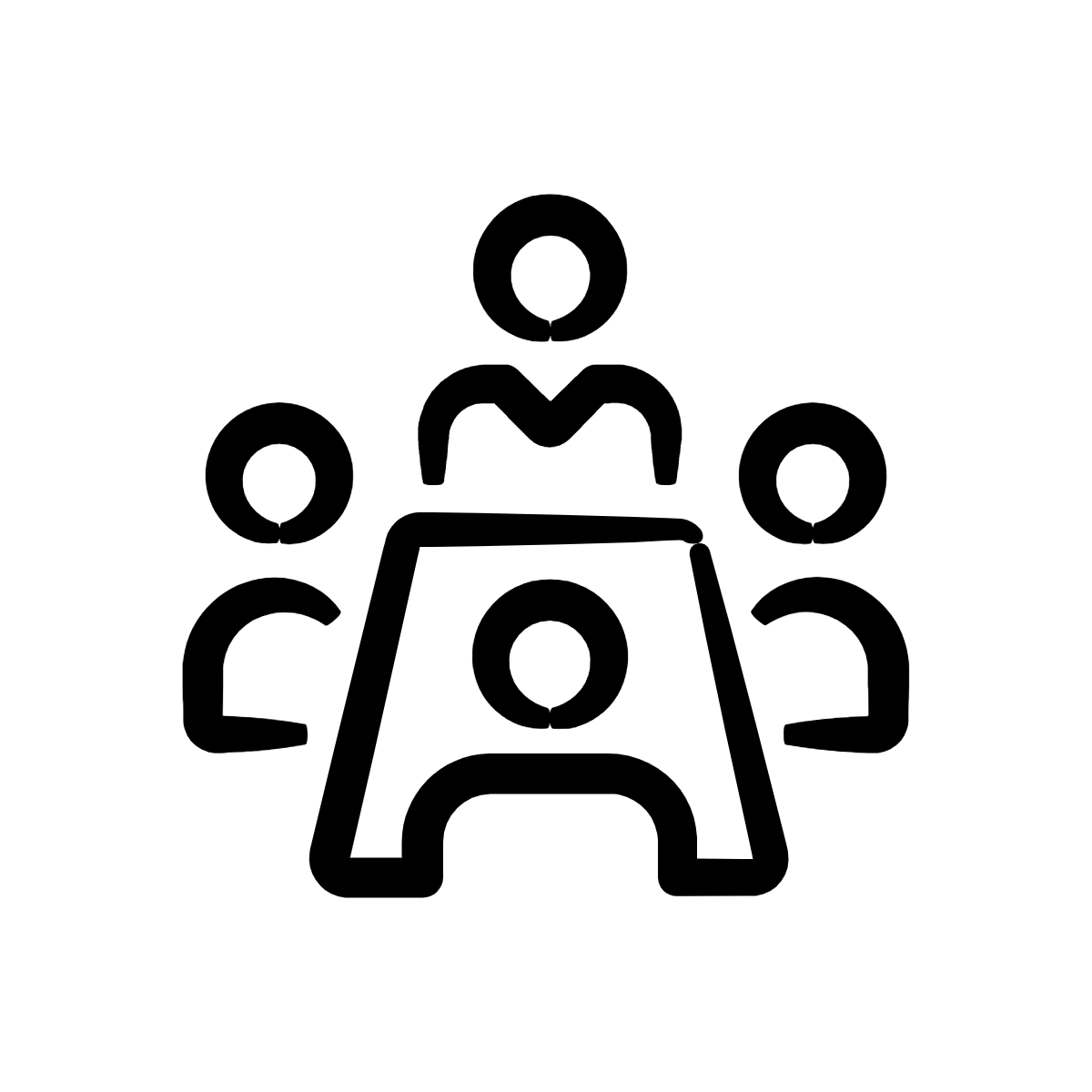}
Meetings were a highly featured work-specific activity. There were three types of meetings were included: status meetings, kick-off meetings, and impromptu meetings. The most common of the three were status report meetings which include daily scrum-style stand-ups or individual check ins with a manager~\citeVideo{M1_29}. These meetings provided an opportunity for developers to ask for help: \inlinequote{Having a stand up in the morning definitely helps with that because it gives you a chance to say 'hey I'm struggling a little bit this one can I get some help? and people are more than happy to jump in}~\citeVideo{M1_34}. Project kick-off meetings provide the opportunity for new project team members \inlinequote{to learn exactly who else is on the team \ldots to not plan work but how we want to work, what kind of meetings we want, how we want to track our work and that kind of thing}~\citeVideo{M1_75}. Videos also showcased team meetings later in the project life cycle where developers collectively planned project sprints at the beginning of a project or conducted retrospective meetings at the end of a project. 

Other types of meetings included unscheduled meetings prompted by a colleague stopping by their desk to clarify a task, or a sudden huddle of co-workers around a developer's code on the screen. This broad variety of meetings showcased the breadth of communication that took place within a team.

\activitysectionnew{Collaboration}{50}{4.30}{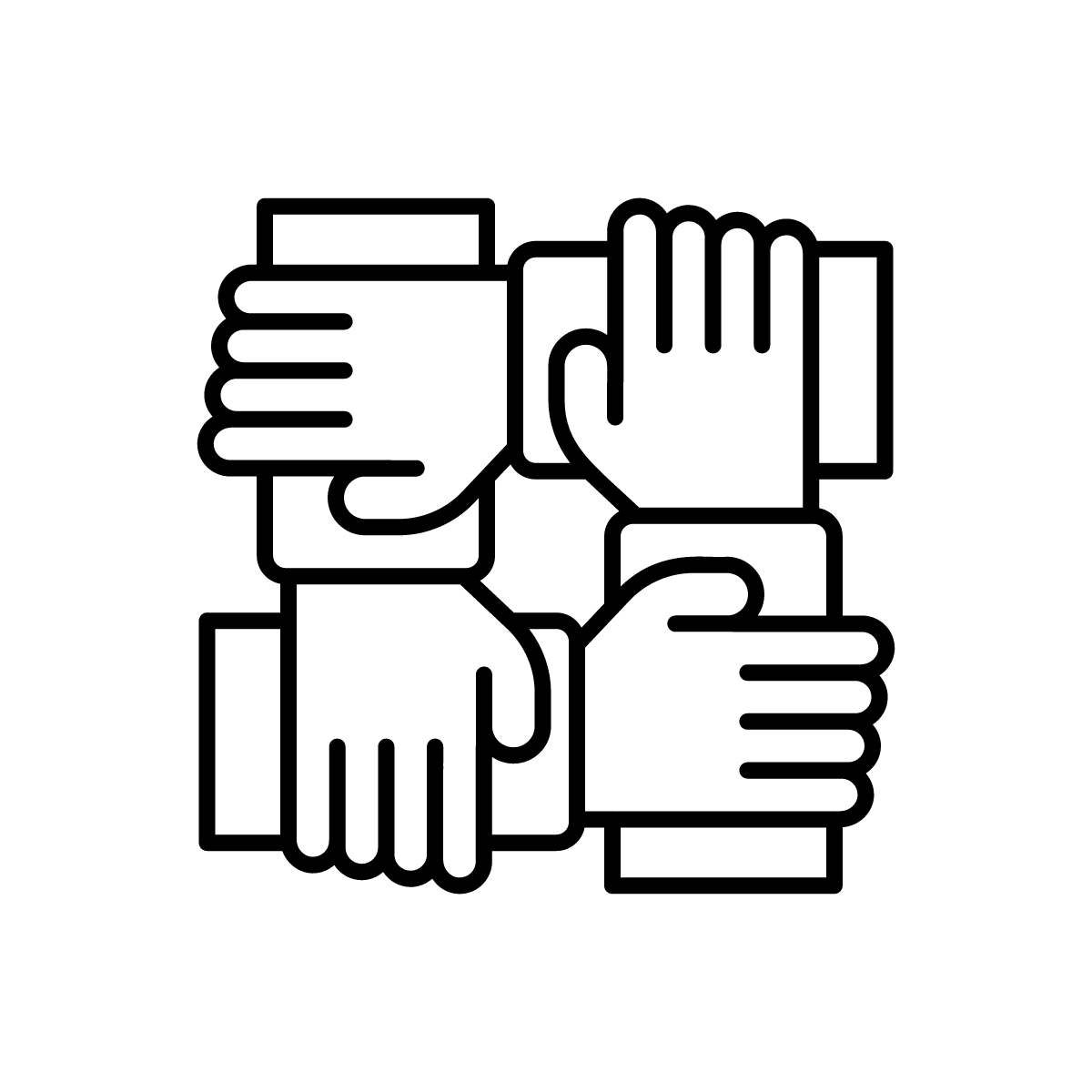}
The collaborative nature of software development was also frequently showcased in a software developer's day. Collaboration in the vlogs went beyond working with others on their immediate team, but also included working with designers, architects, marketing as well as other developers from different products. Videos included colleagues from different roles collaborating at the whiteboard where developer spent a considerable amount of time in \inlinequote{cross team collaboration and engagement because as a team [we] are not exactly as an island. So whatever the solution we are building and designing should be seamless with other systems} \citeVideo{M1_48}.

Likewise, collaborations also occurred asynchronously through emails and communication tools like Slack~\cite{zhang2018making}. Most vlogs showed developers beginning their workday by \inlinequote{looking at emails for both work and quickly for other tech stuff
}\citeVideo{M1_04}. For developers in companies with teams in different countries \inlinequote{a lot of emails come in from Seattle or Japan or other places of the world and they're coming overnight. So every day every morning I get through a long list of things that happened, new announcements, new information, all the system alerts, the information about the metrics}\citeVideo{M1_08}. Similarly, software developers who were freelancers found themselves collaborating with other developers in share co-working spaces.
In essence, regardless of affiliation to a large tech company or being a independent freelancer, developers working remotely emphasized the importance of communication. Feeling connected with their peers and clients or other stakeholders is valuable for developers to ensure meaningful progress.

\activitysectionnew{Learning}{12}{2.0}{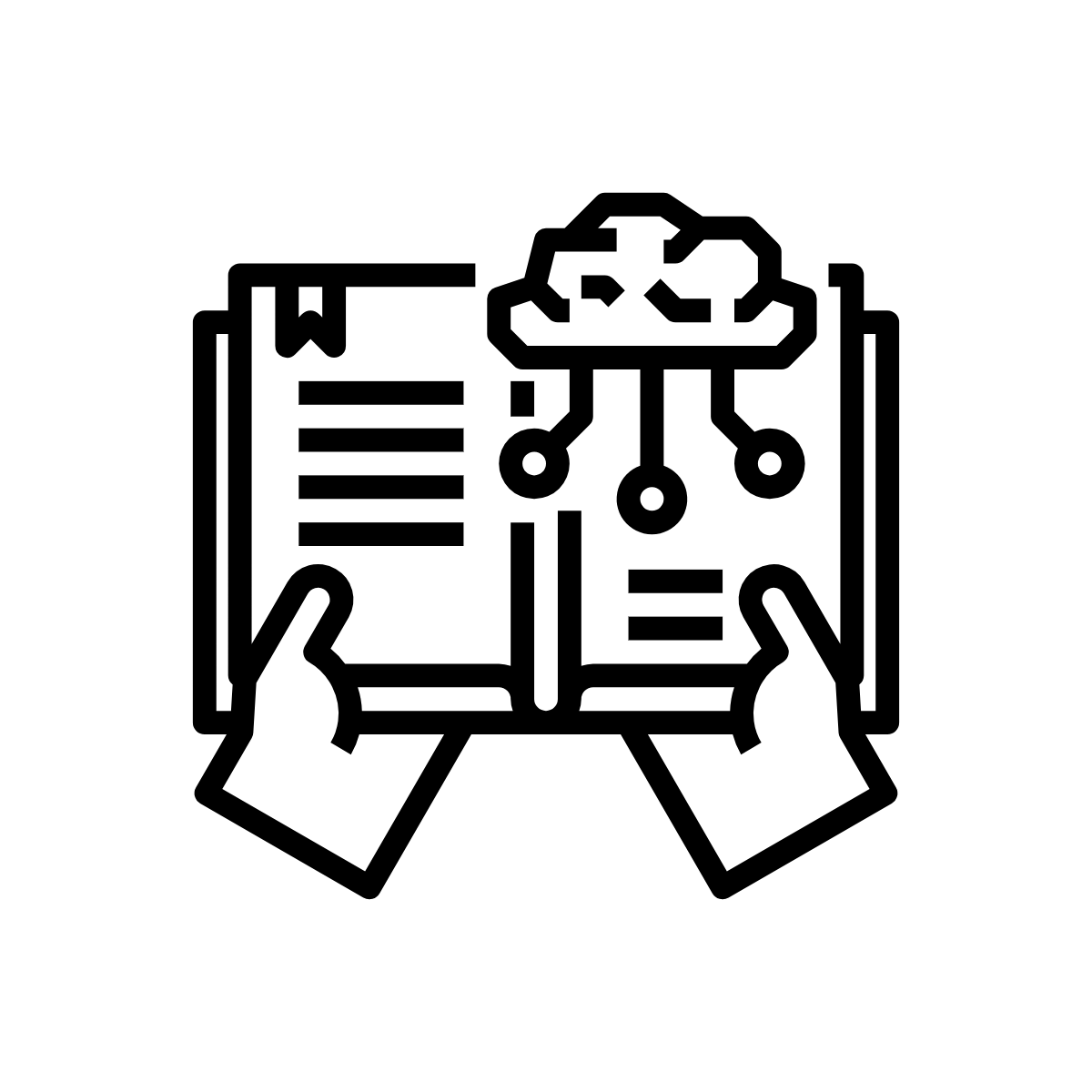}
Videos also featured developers talking about the different approaches how they learn new technical concepts and refresh themselves on old ones. For example, videos included learning new programming languages like JavaScript~\citeVideo{M3_138} or Redux~\citeVideo{MS_9}~\cite{redux} and getting familiar with new machine learning paradigms. Developers described their undergoing various training exercises that last up to 3 hours once or twice a week, and can be in the form of assignments or hands-on workshops or refresher training at the end of the workday \citeVideo{M1_11}, \citeVideo{M2_103}, \citeVideo{M5_152}.
This requirement of continuous learning was also especially important for freelancers. As \lookupVideo{M1_22} explains in freelance development \inlinequote{you switch code bases quite often and there's always a ramp up time when you're learning a new code base trying to identify the issues that you have been hired to solve.}
Likewise,  self-taught programmers referenced using only coding platforms such as CodeAcademy~\cite{codeacademy} and Udemy~\cite{udemy} as alternatives to a traditional Compute Science degree from a university. Developers in videos also described how non-traditional course help facilitate their career transition from  being a creative writer or music producer to a being a software developer.

\activitysectionnew{Hobbies and Side Projects}{35}{5.10}{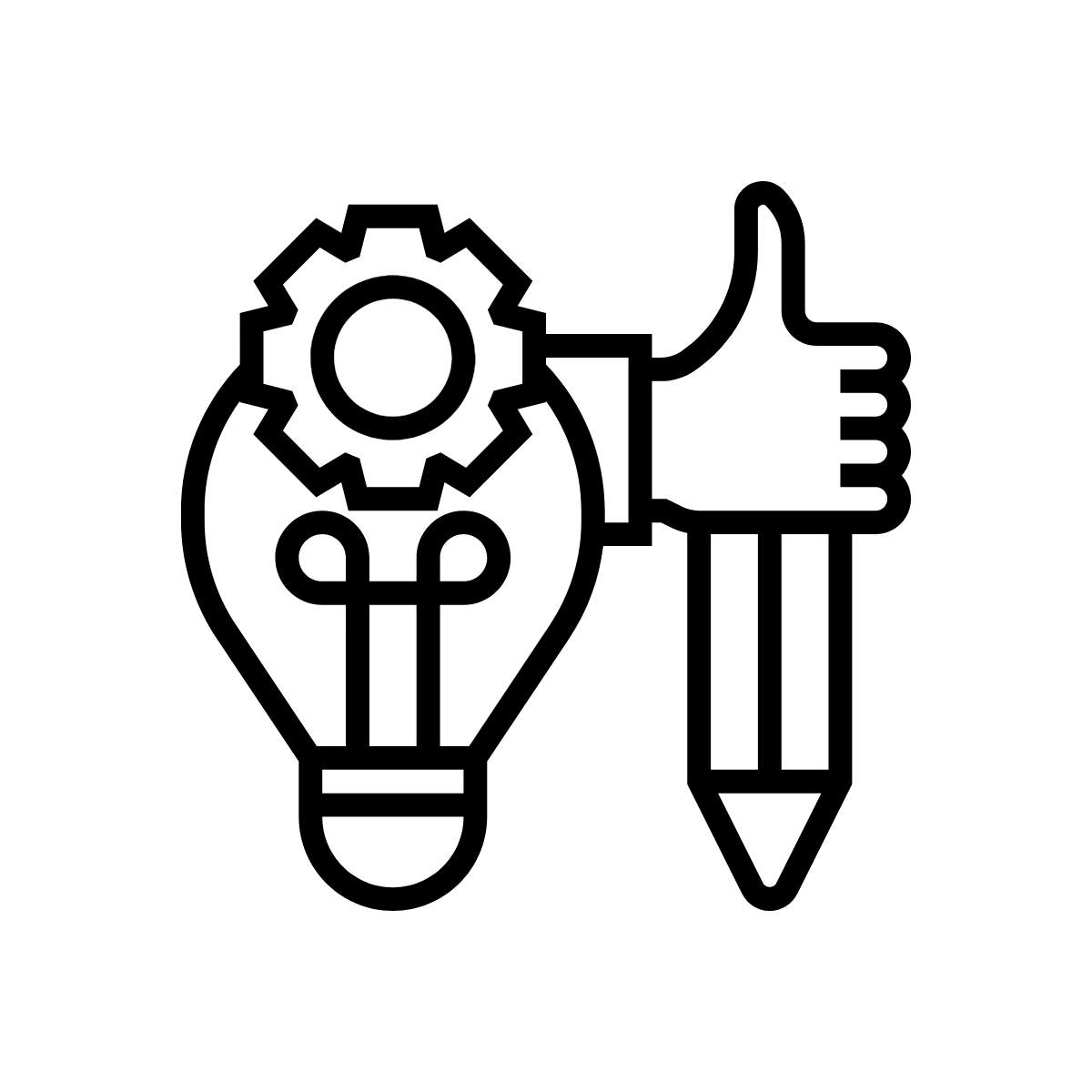}
Developers and engineers talk about various creative hobbies they engage in after work hours. After-work activities include social activities like participating in beauty pageants and networking with various PR managers\citeVideo{M1_06}, volunteering at local events organized by Youth At Risk programs \citeVideo{M2_101}. Others take the time to work a second job like teaching at a computer science institute in Philippines \citeVideo{M1_49}.
Some other developers enjoying creative hobbies like baking for a self-run patisserie \citeVideo{M1_27}, ``as a photographer and a videographer" editing and taking photos and videos for events like engagement photo shoot \citeVideo{M1_17}. 

Developers who upload content on \yt have to spend additional time editing videos, recording and editing voice overs, and uploading videos on \yt. Some developers who vlog also have large number of followers on other social media platforms like Instagram~\cite{instagram} and they invest in creating graphic design flyers and editing images about programming, software development, web development etc. 

Developers also code in alternate platforms as a creative endeavor in free time like making applications using SwiftUI~\cite{swiftui} \citeVideo{M1_27}, \citeVideo{MS_2}, or creating games on Unity Engine~\cite{unity} \citeVideo{M1_38}, \citeVideo{M1_57} and releasing them on App Stores and gaming platforms like Steam~\cite{steam}. Some developers take their passion projects a notch higher and turn them into start-ups [\lookupVideo{MS_7}, \lookupVideo{M1_60}] and spend after work hours collaboration, planning, and coding.

\activitysectionnew{Social Activities:}{37}{8.1}{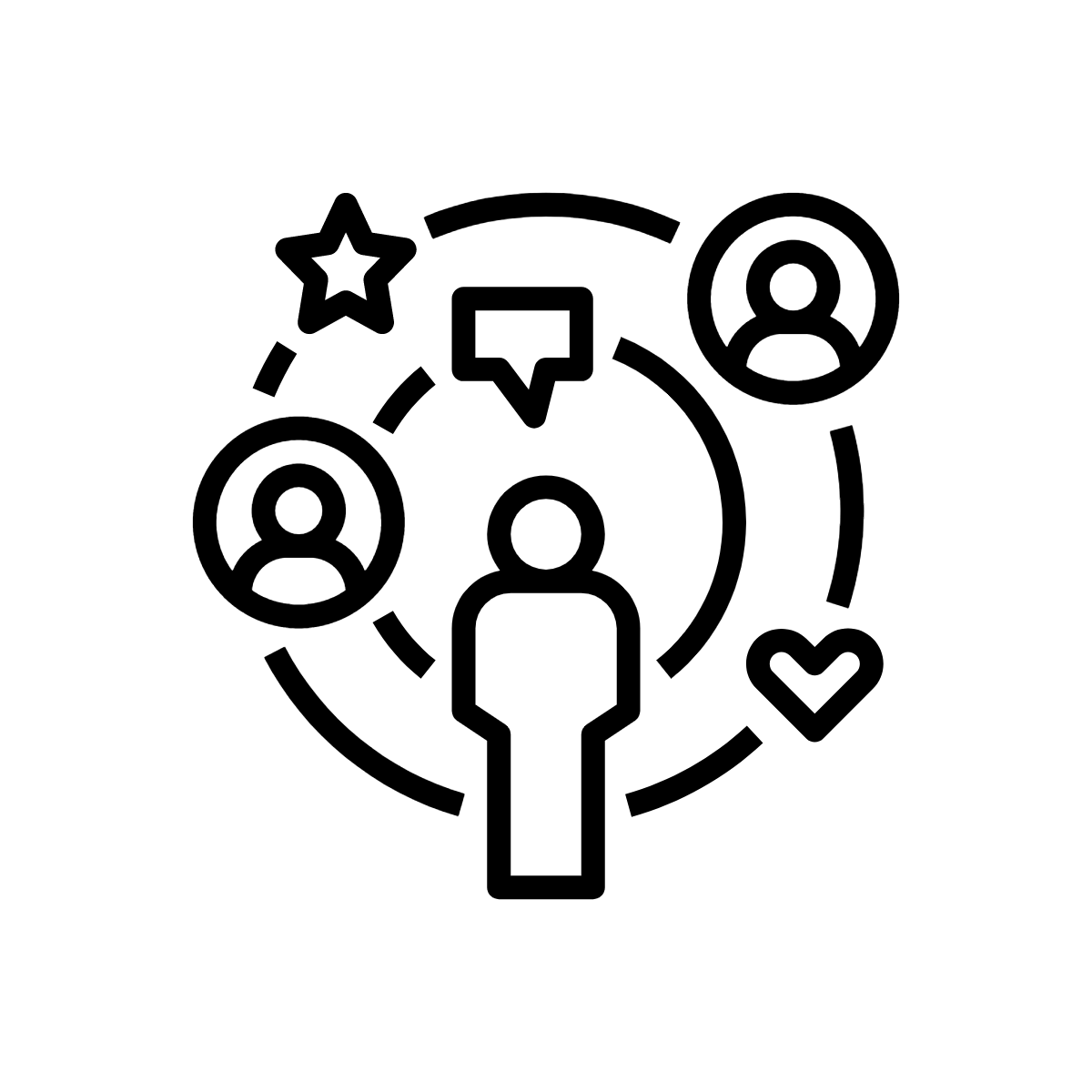} 
Outside work and hobbies, developers spend their time with family and friends and discuss the importance of weaving in social time to prevent burnout. Some developers maintain a strict schedule of work and make sure to take free hours between 6-9 in the evening and ``do whatever [I] want, watch TV, play games  in the evening, whereas other developers show the ability to have a flexible schedule and balance work-intensive days when they leave work late at night and shorter work days for special plans like anniversaries \citeVideo{M2_124} or trips.
Vlogs show developers spending time with friends through various sports (like golfing, basketball, and martial arts) and video games [\lookupVideo{M1_02}, \lookupVideo{M2_99}, \lookupVideo{M2_133}, \lookupVideo{M3_135}, \lookupVideo{M5_149}], going out to restaurants and bars \citeVideo{M1_13} and movies \citeVideo{M2_133}, playing trivia \citeVideo{M1_93}, video games with friends and family  or spending time with kids after their babysitter leaves \citeVideo{M1_22}.

\activitysectionnew{Health, Workout, and Lifestyle:}{100}{47.3}{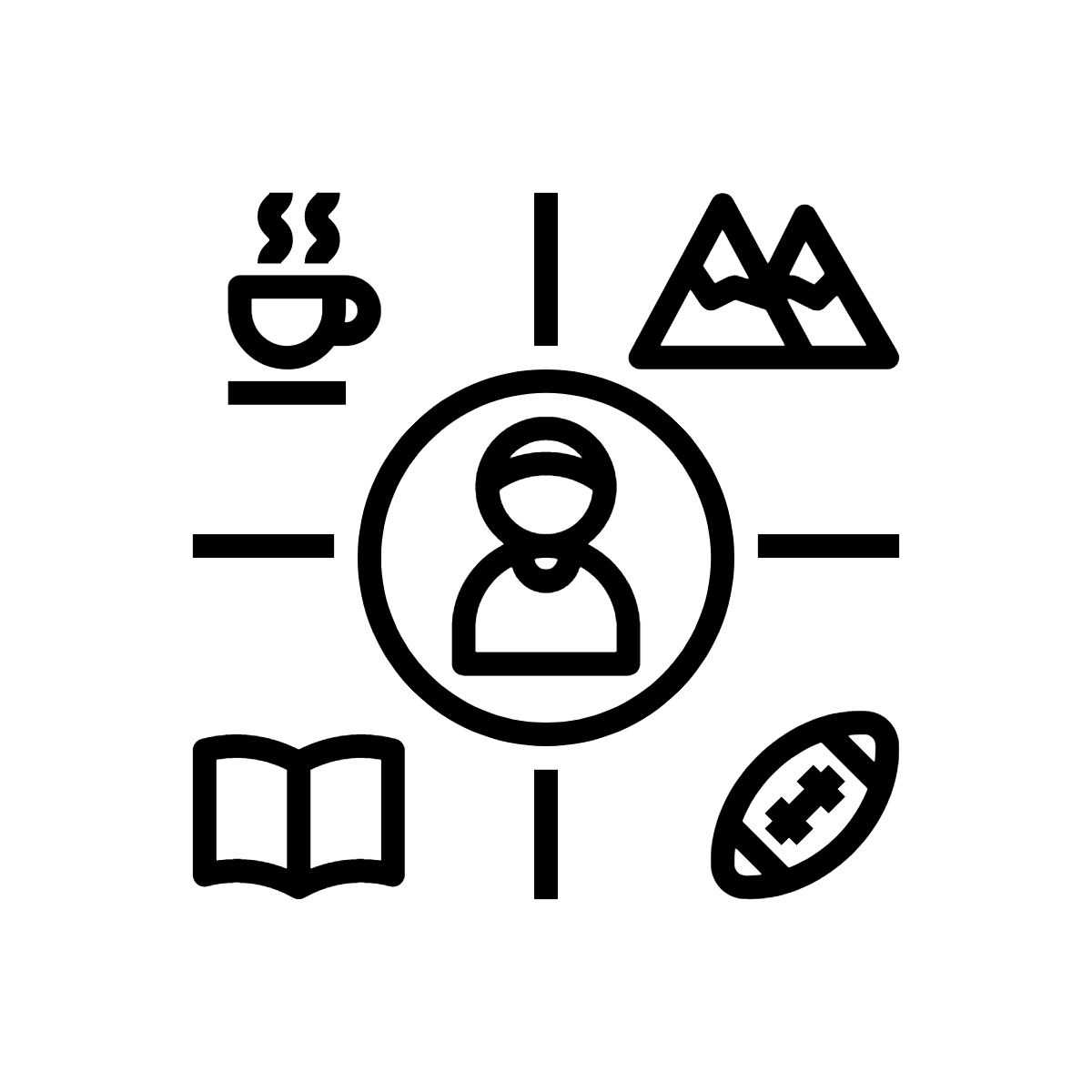}
A large portion of the vlogs show various lifestyle related activities like workouts [\lookupVideo{M1_31}, \lookupVideo{M2_115}, healthy food recipes and eating habits. Food is an essential part of lifestyle and health, and 89 vlogs included various unique cuisines, food they can afford and enjoy at home and lunch spreads and buffet at work [\lookupVideo{M1_05}, \lookupVideo{M4_145}, \lookupVideo{MS_18}].

Developers emphasize on weaving in some from of exercise, from lightweight exercises like stretching \citeVideo{M1_22} or a walk \citeVideo{M5_153}, to indoor training \citeVideo{M2_95} and intensive workouts at the gym [\lookupVideo{M2_133}, \lookupVideo{M4_146}]. Some others train and practice areobics or various sports as a form active workouts like boxing \citeVideo{M1_09}, basketball \citeVideo{M1_14} or going on a long run \citeVideo{M1_65}.

Developers talk about physical health concerns they have and are commonly discussed among their co-workers like chronic back pain \citeVideo{M2_112}, and also discuss various measures to take care of these health issues like using a wearable posture corrector \citeVideo{M3_138}, frequently switching between sitting and standing using stand-up desks \citeVideo{M2_119}, using Pomodoro techniques to take stretching breaks \citeVideo{M2_102}. Mental health concerns also features in the vlogs. \inlinequote{Sitting long hours in front of the desks causes tension headaches, stress} \citeVideo{M1_39} especially with \inlinequote{schedules that look like dumpster fires} \citeVideo{M1_93}. Developers discuss different small but effective self-care habits they use to counter the stress like ``hydrating in the morning before workout'' \citeVideo{M1_57}, meditation \citeVideo{M1_42}, taking protein shakes and supplements \citeVideo{M5_149}, taking time off and going on trips [\lookupVideo{M1_20}, \lookupVideo{M1_33}].

Beyond these, vlogs also portray the kind of lifestyle developers lead including the types of apartments and cars they can afford [\lookupVideo{M1_20}, \lookupVideo{M2_131}, \lookupVideo{M1_93}], the gadgets and equipment they use [\lookupVideo{M1_22}, \lookupVideo{M1_55}] or their work space setup at home, and the physical work environments (like the campus and office facilities, conference rooms, cafes \& lounges, and kitchen [\lookupVideo{M1_25}, \lookupVideo{M1_69}, \lookupVideo{M2_95}]. While some vlogs show the luxurious lifestyles of living in a penthouse, traveling and driving expensive cars, other vlogs focus on the ``regularity'' of the lifestyle which includes home made meals \citeVideo{M3_137} or doing daily house chores \citeVideo{M2_125}.

\smallskip

\smallskip
\mybox{\textbf{Summary:} About half of the vlog-time is dedicates to show activities developers perform throughout the day, most prominently coding and peripheral coding tasks like debugging, meetings, and administrative tasks.

\vspace{4pt}
Apart from work, a large portion of a developer's day is presented as time spent on lifestyle, health and workout (47.3\% of the vlog durations) and doing social activities (8.1\% of the vlog durations). Developers also presented themselves doing creative work quite often (5.1\% of the vlog times).}

Our distribution of mean time spent varies from existing literature as it also captures the time spent beyond work through activities like health and socio-personal life. This look beyond just work time puts a different perspective on ``who is a developer''--- or at least how they work.

\subsection{\RQB (RQ2)}
\label{sec:result-rq2}

\begin{figure}
\includegraphics[width=\linewidth]{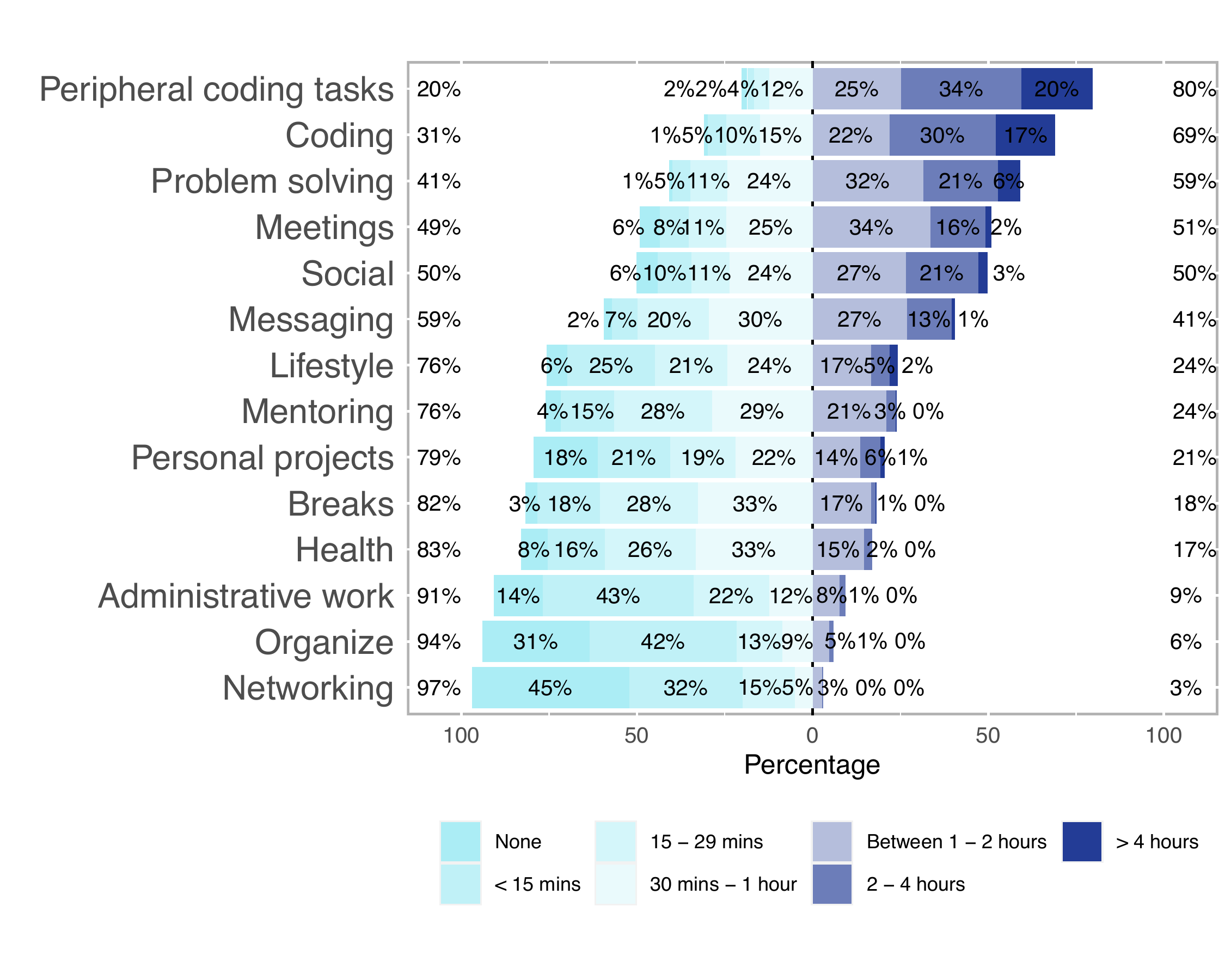}

\caption{Distribution of number of developers and their (interval of) time spent in each activity. Time spent is measured from none (left bars, darkest) to $>$4 hours (right bars, darkest). Left margin percentages show developers who chose $\leq$1 hour on an activity (light blue bars) and right margin percentages show those who spent $>$ 1 hour.}
\label{fig:selftime_distribution}
\end{figure}

Figure~\ref{fig:selftime_distribution} presents the distribution of time spent in the 14 activities as reported by survey respondents.
The x-axis reflects the percentage of respondents across time intervals and the y-axis reflects the 14 activities. The diverging stacked bar chart splits time spent into how survey respondents spent less than an hour in an activity (left side of plot) or more than an hour on an activity (right side of plot).
Next, we highlight the activities with the most and least time spent as well as additional findings from this analysis.

\smallskip
\emph{Most Time Spent.} Among all activities, industry developers spent the most time on activities directly related to work - coding (69\%), coding related work (debugging, testing etc.) (80\%), and problem solving (51\%). 

\smallskip
\emph{Least Time Spent.} Industry developers spent the least amount of time (less than an hour) networking (97\%), organizing their work day (94\%), and performing administrative tasks (91\%).

\smallskip
\emph{Outside of Coding.} We found an interesting distributions. When it came to life outside of work. For instance, for how much time respondents reported spending with friends and family we found an even split: 50\% reported spending less than an hour and the remaining 50\% spending more than 1 hour. Likewise, for personal projects (21\%), lifestyle (24\%), breaks (18\%), and maintaining health (17\%) we found that very few developers were able to devote more than hour to these activities.

\smallskip
\emph{Enhanced Collaboration.} We also find that developers often spent time in collaborative activities like 51\% developers spending more than an hour a day in meetings and 41\% on messaging. We should note that as the survey was conducted in the peak of the COVID-19 pandemic this could influence their collaboration style. Developers were mostly working from home, which might affect the time spent in video call meetings and messages they exchanged.

\smallskip
\mybox{\textbf{Summary:} Developers from our survey  largely spend their time working on technical aspects (coding or peripheral coding tasks) of their work. However, when not working on code, their time is spent in meetings and staying connected, or mentoring co-workers. 

\vspace{4pt}
With little time left in the day, less than 20\% developers spend time on their physical or mental health, taking breaks, or to plan out their day.}

\subsection{\RQC (RQ3)}
\label{sec:result-rq3}

\begin{figure}
\includegraphics[width=\linewidth]{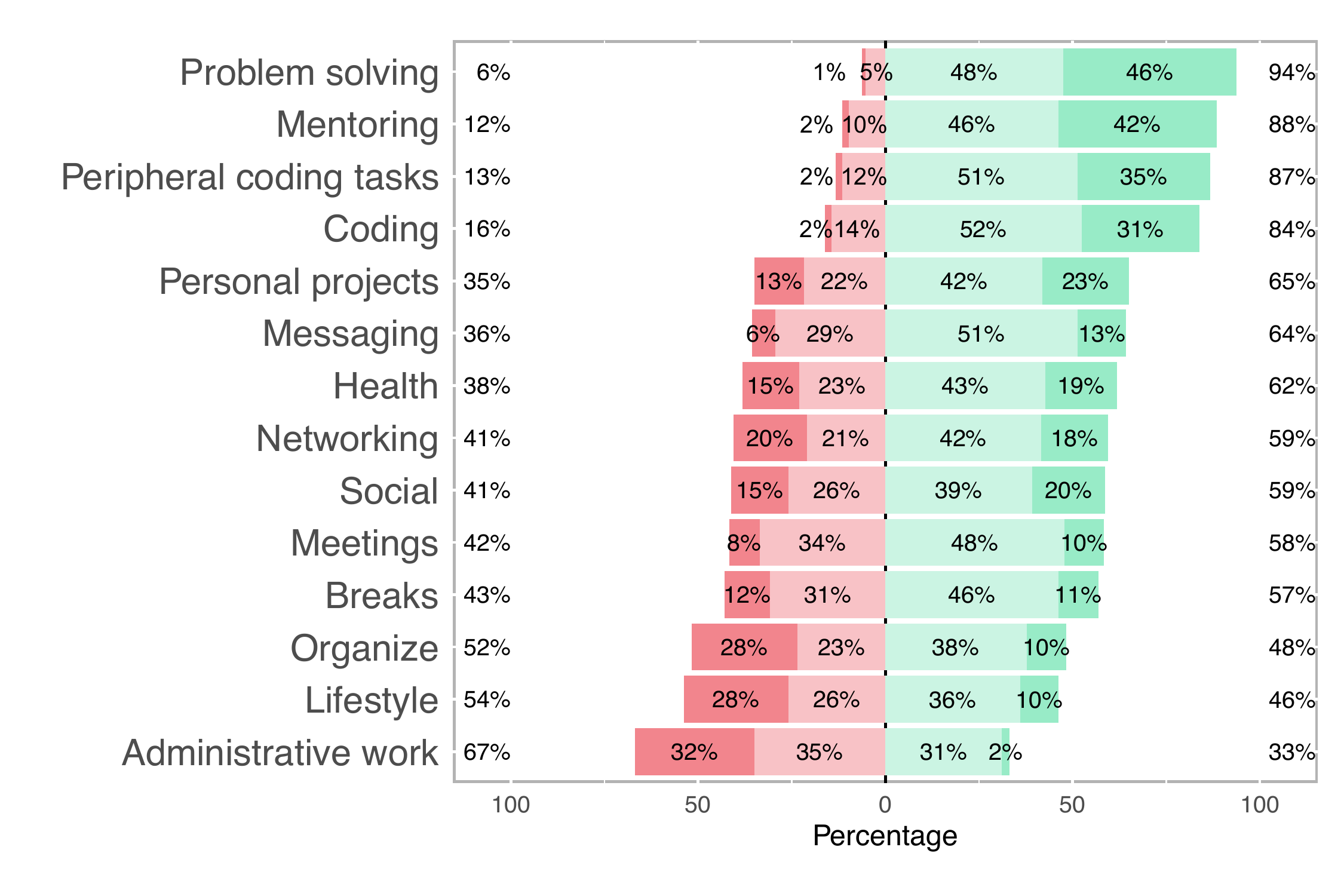}
\includegraphics[width=\linewidth]{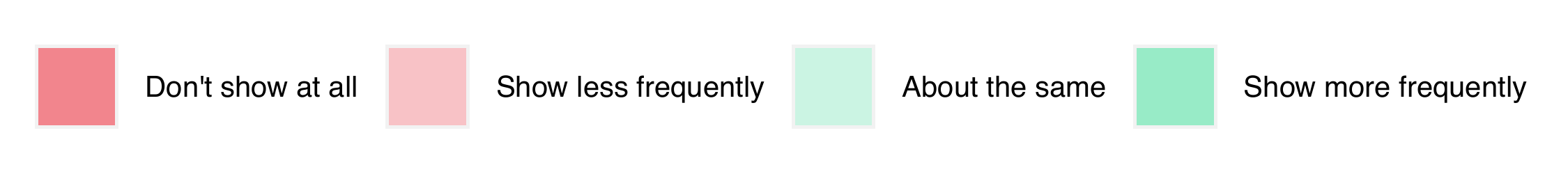}

\caption{Distribution of screen-time per activity developers would dedicate in \pretendvlogs compared to the reported time they spend on each activity; developers can choose between `Don't show at all', `Show less frequently', `About the same', and `Show more frequently'. The left margin shows percentage of developers represented by the red bars, and the right margin shows those of the green bars.}
\label{fig:all_self_vlog}
\end{figure}

To understand what developers find valuable to share as part of their ``day in life" identity, in the survey we asked \MSdevs to make \pretendvlogs (refer to section~\ref{sec:methodology:survey} for specific question). Figure~\ref{fig:all_self_vlog} shows a distribution of what activities the developers would dedicate `more' or `less' vlog-time in their \pretendvlogs compared to their reported-time.

Most developers (94\%) find value in showing their problem-solving process through \pretendvlogs similar or more than their reported time. 88\% developers also think its valuable to showing mentoring activities in their \pretendvlogs.

Since core technical activities (coding and code peripheral) takes up most of their reported time, developers want to capture them in \pretendvlogs (refer to Figure~\ref{fig:selftime_distribution}). 87\% and 84\% developer (respectively) want to dedicate similar or more vlog-time showing parts of their daily life spent in coding-related activities and coding. Additionally, 65\% developers want to capture their time spent on personal projects. Other than technical activities, 64\% developers want their \pretendvlogs to feature time spent in messaging co-workers and staying connected at work, networking(59\%), and meetings at work (58\%). They also want to emphasize the breaks they take in between work (57\% developers).

Outside work, 62\% developers find it worthwhile to include health related activities in their \pretendvlogs, and 59\% developers want to capture their daily social interactions.

However, the activities developers want to capture more (or similar) in their \pretendvlogs are different from the activities they reported spending more time on (see Figure~\ref{fig:selftime_distribution}). This can stem from the motivations behind the vlogs. To understand how motivation can affect the value of presenting an activity though vlogs, we randomly assigned a motivation scenario to the survey question ``If you were to make a vlog \ldots". 112 developers were asked to promote personal network and value, 106 developers were to promote diversity in computing, and 117 developers to promote awareness of computing careers through their \pretendvlogs.

\begin{table}

\newcommand{\babypenguin}{\cellcolor{green!25}}

\caption{The percentage of developers who would allot ``more" screen-time to an activity compared to the real-time spent. For each column (motivation), green colored cells mark activities at least 20\% developers want to allot ``more" screen-time. Motivations for which an activity is statistically different from the others motivations are labelled with (*).} \label{tab:motivations_compare}
\small

\begin{tabular}{lccc}
\toprule
Activity & Personal brand & Diversity & Awareness \\
\midrule 
Coding & \babypenguin 32\% \phantom{(*)} & \babypenguin 29\% \phantom{(*)} & \babypenguin 35\% \phantom{(*)} \\ 
Peripheral coding tasks  & \babypenguin 36\% \phantom{(*)} & \babypenguin 35\% \phantom{(*)} & \babypenguin 36\% \phantom{(*)} \\ 
Problem solving & \babypenguin 43\% \phantom{(*)} & \babypenguin 50\% \phantom{(*)} & \babypenguin 46\% \phantom{(*)} \\ 
Administrative work &  2\% \phantom{(*)} &  1\% \phantom{(*)} &  4\% \phantom{(*)} \\ 
Mentoring & \babypenguin 41\% \phantom{(*)} & \babypenguin 56\% (*) & \babypenguin 33\% (*) \\ 
Organize &  12\% \phantom{(*)} &  12\% \phantom{(*)} &  8\% \phantom{(*)} \\ 
Meetings &  6\% \phantom{(*)} &  17\% (*) &  9\% \phantom{(*)} \\ 
Networking &  11\% \phantom{(*)} & \babypenguin 26\% (*) &  17\% \phantom{(*)} \\ 
Messaging &  9\% \phantom{(*)} &  17\% \phantom{(*)} &  11\% \phantom{(*)} \\ 
Personal projects & \babypenguin 23\% \phantom{(*)} & \babypenguin 30\% \phantom{(*)} &  16\% (*) \\ 
Social &  16\% \phantom{(*)} & \babypenguin 27\% \phantom{(*)} &  17\% \phantom{(*)} \\ 
Health &  14\% \phantom{(*)} & \babypenguin 22\% \phantom{(*)} & \babypenguin 21\% \phantom{(*)} \\ 
Lifestyle &  11\% \phantom{(*)} &  7\% \phantom{(*)} &  10\% \phantom{(*)} \\ 
Breaks &  6\% \phantom{(*)} &  18\% (*) &  9\% \phantom{(*)} \\ 

\bottomrule
\end{tabular}

\end{table}

To understand how motivations affect what activities developers find valuable, we calculated the percentage of developers who chose to dedicate ``more'' screen-time compared to their real-time spent per activity. Since we already know the real-time developers spend on these activities, looking at distribution of developers wanting to capture ``more" captures the relative value developers unconsciously assign when presenting themselves through vlogs.

Table~\ref{tab:motivations_compare} presents the percentage distribution of developers. Developers find more activities valuable when they want to make a vlog to promote diversity, and similar activities as valuable when making vlog to promote awareness about computing careers or promoting their personal ventures.

\paragraph{When motivated to promote personal value and network.} $>30\%$ of developers want to show coding, peripheral coding related and problem solving activities through the \pretendvlogs; most developers ($43\%$) want to dedicate more vlog-time for problem solving compared to their reported time. Mock-up vlogs will have higher dedicated time for mentoring and personal projects as $41\%$ and $23\%$ developers want to show these activities more compared to their daily life.

\paragraph{When motivated to promote diversity in computing.} Developers find a lot of activities valuable when motivated to promote diversity. Mentoring is an important activity to show in these \pretendvlogs with the 56\% developers interested in including it, followed by problem solving (50\% developers). Other than activities at work, developers also find creative personal projects (personal project ($30\%$) and Networking ($26\%$)), social life (friends and family ($27\%$)), and health related activities ($22\%$) important activities to include in the \pretendvlog.

\paragraph{When motivated to promote awareness about computing careers.} Most developers want to capture problem solving (46\%), coding peripheral (36\%), and coding (35\%) activities to in \pretendvlogs. 33\% developers want dedicating more vlog-time on mentoring, and 21\% developers want to show health related activities more compared to their reported-time.

\smallskip
\mybox{\textbf{Summary:} When asked about making a vlog, most developers find  coding, coding related activities, showing problem solving, and engaging in mentoring activities valuable and want to dedicate more screen-time for these activities in the vlogs. 

\vspace{4pt}
However, motivation behind making the vlogs affects what other activities are deemed valuable; when promoting personal brand and growth developers want to capture their time working on personal projects in the vlogs, when promoting diversity developers find it valuable to show socio-personal sides of their life, and when promoting awareness about computing careers developers want to feature their time spent in maintaining a healthy life.}

\section{Discussion}
\label{sec:discussion}

Our findings have brought together two perspectives of software developers.
From \textbf{RQ1}, we presented an expanded understanding of the types of activities developers to in a day. Likewise, from \textbf{RQ2} and \textbf{RQ3} we found that similar activities would be of interest in presenting to other developers (who do not vlog) albeit at different frequencies.
In this remainder of this section, we interpret our findings from our research questions to help build a more complete understanding developers identity.

\label{sec:discussion:picture-perfect}

In answering our research questions we found that \MSdevs wanted to their vlogs to paint a \emph{not so perfect picture} of what they do in practice. Next, we discuss interesting differences in how they reported spending their time in contrast with what developers wanted to display publicly in their \pretendvlog.  Table~\ref{tab:contrast} compares the findings from Section~\ref{sec:result-rq2} with Section~\ref{sec:result-rq3}.

\subsection{Picture (Not So) Perfect: Contrasting Reported Time vs. Mock-Up Vlog Time}

\paragraph{Presenting activities that take significant time.}
\MSdevs wanted their \pretendvlogs to capture the activities they also spend the most time on---\emph{coding, peripheral coding activities,} and \emph{problem solving}. Between 59\% and 80\% of developers reported they spend an hour or more in these activities and between 84\% and 94\% of developers wanted to include them in the \pretendvlogs.
Developers also wanted to capture how they communicate and stay connected with their colleagues: 41\% developers spend an hour or more in emails and different messaging services and 64\% developers want their \pretendvlog to appropriately include that activity. %

\paragraph{Presenting more than in real life.}
\MSdevs want their \pretendvlog to show some activities \textbf{more frequently} even though they \textbf{reported spending less time} on these activities. While less than 25\% developers reported spending an hour on a typical day \emph{mentoring} others, 88\% developers want their \pretendvlog to feature mentoring with 42\% of them want to dedicate more time in the vlog showing mentoring in proportion to their actual time spent. We found a comparable difference with respect to \wellnessactivities. About 80\% developers reported spending less than an hour on \emph{health activities, lifestyle,} and creative \emph{personal projects} outside of work. Yet many respondents aspired for their \pretendvlog to present these activities to showcase a healthy lifestyle.%

\paragraph{Presenting less than in real life.}
We found that activities which \MSdevs wanted to disclose less on their \pretendvlog than they do in their life. One of these activities is attending \emph{meetings}: 51\% developers spent more than an hour a day doing meetings, yet 42\% developers want their \pretendvlog to show comparatively less. Meetings are unavoidable at work, ``and some days [developers] spend the whole day in meeting rooms''~\citeVideo{M47}. In any setting, in-person or virtual, a day full of meetings can cause fatigue (recently zoom fatigue). Meetings are a necessary evil, that developers did not want to define as a major part of their job as a developer. 
Similarly, developers reported spending more \emph{social} time with friends and family than what they would portray in their \pretendvlog. This may be due to privacy concerns impacting how much \MSdevs are comfortable to share about the people around them. 

\paragraph{Differences by \pretendvlog motivation.}
The prescribed motivations to vlog impacted what developers wanted their \pretendvlog to show. To  \emph{encourage diversity}, developers wanted their \pretendvlog to capture the technical and social aspects of a work day. Respondents crafted a day that incorporates family time, ample breaks, shares collaboration through meetings, building a network, and empowering others through mentoring. This large variety of activities (showing $>$20\% in Figure~\ref{fig:selftime_distribution}) indicates that in order to share diversity of the development experience, they should share a variety of activities. %
When the motivation was to \emph{encourage awareness} about computing careers, developers wanted their \pretendvlog to spend less time on personal projects compared to other motivations. One reason could be that developers want to transparently let their audience know they will have less time to devote to their computing hobbies and inherently more time that will be spent elsewhere.
\begin{table}
\newcommand{\babypenguin}{\cellcolor{red!25}}
\newcommand{\doublerainbow}{\cellcolor{green!25}}

\caption{Contrasting time spent with \pretendvlogs. The ``High'' column has activities where at least 40\% of survey respondents reported that they spend one hour or more. The ``More'' row  has the top half of activities ranked based on how frequently respondents wanted to show them in the \pretendvlogs.}
\label{tab:contrast}
\small 
\begin{tabular}{p{0.20\columnwidth}p{0.15\columnwidth}p{0.25\columnwidth}p{0.23\columnwidth}}
\toprule
 & & \multicolumn{2}{p{0.45\columnwidth}}{\centering \bf Reported time spent} \\[2pt]
 & & \doublerainbow High & \babypenguin Low \\
 \cmidrule{3-4}
\multirow[t]{3}{=}{\bf Featured in \pretendvlogs} & \doublerainbow More & Problem solving, \newline Coding, Peripheral coding tasks, \newline Messaging & Mentoring, Health, \newline Personal Projects \\
& \\[-8pt] %
& \babypenguin Less & Social, \newline Meetings & Lifestyle, Breaks, \newline Administrative work, Organize, \newline Networking\\ 
\bottomrule
\end{tabular}
\end{table}

\medskip
Many of the differences we find across reported time and \pretendvlog time may be based on what developers anticipate their day should be in comparison to what it in fact entails. Further research could investigate if these differences persist in other developer ``day planning'' studies.

\subsection{Vlogs for Future Developer Studies}
Through our study, we found that vlogs can be used as a dataset for understanding developer experiences with several advantages. We outline a few opportunities below.

\smallskip
\textbf{Vlogs can be used as a data source to understand self presentation}, which typically takes on two forms: self-constructive or audience-pleasing~\cite{baumeister1987selfpresentation}.
A \emph{self-constructive} vlog emphasizes aspects that developers themselves find valuable, and captures activities that developers enjoy and learn from, and identify with.
The vlogs created for \emph{audience-pleasing} aspects tend to serve the audience~\cite{jensen2003selfpresentation} through information that the audience will find useful. One instance of this is when the audience consists of other developers looking for a career or location change as well as students contemplating a career in computing~\cite{cummings2019vlog}. Developers sharing vlogs along these lines will share what they think is be most valuable to know when one is going on the job market (self-constructive) and hidden secrets to success on the job market that the audience may not know (audience-pleasing). 
These two forms make vlogs a prosperous source of readily available data from developers around the world, and can be used to understand how the information that developers share through vlogs affect developer communities? How can vlogs be used to increase diversity in computing? When triangulated with other sources, research can also draw valuable insights from vlogs like (3) how have workplace practices and types of activities evolved over the years? Eventually, vlogs evolve into an alternative form of documentation with the ability to directly link to source code.

\smallskip
\textbf{Vlogs can be used as a time capsule} for understanding developer activities over time. Vlogs and other types of videos can capture trends in tech companies, trends in programming languages, and even developers fashion choices (e.g., life before developers wore hoodies) to study them over time periods. 
One example of a time period that can be studied is the difference between collaboration challenges developer faced during the COVID-19 pandemic compared to challenges the pandemic---a unique time period where many software developers were working remotely. 
Using vlogs as a time capsule provides the opportunity for studies to even compare developers at the inception of video sharing~\cite{zoo} to the present.

\smallskip
\textbf{Vlogs can be used to study the diversity of software developers.} 
When studying diversity, small groups of developers (e.g., Trans*, Black, visually-impaired, etc.) may be over-sampled for empirical studies due to the small size of their population, leading to study fatigue.
Vlogs can be the first dataset researchers use for an analysis of these group's behavior and expertise before asking the same group of developers to take another survey, thus reducing survey fatigue. 
Likewise, if there are not already vlogs for this group, researchers could design a study for developers in this study to share details about multiple parts of their experience to use as a reference for many studies to come.
Another example of vlogs being a resource to reach out to demographics of developers we traditionally don't have access to such as developers across geographic regions (e.g., a researcher in the Netherlands studying developers in Japan). The breadth of vlogs provides a snapshot at the intersections between technical and socio-cultural experiences around the world. 

\medskip
Future researchers should be mindful of the opportunities (and challenges) that this type of developer populations present with respect to their studies.
We acknowledge that vlogs are not a substitute for conducting other types of empirical studies of developers, however, we hope researchers refer to them as an alternative for conducting empirical studies.

\section{Limitations}

We used a mixed method study approach based on vlogs on \yt and a survey in a large software company.
Findings from qualitative research are difficult quantify, so we triangulate our insights from the vlog analysis with a survey of 335 developers. As with any empirical study, there are limitations to this work.

For the \emph{vlog data}, the videos may not be an accurate depiction of an actual workday. For example, to stand out developers may vlog stories that are unique and interesting, add elements like humor and suspetnse, and adhere to current trends like work-life balance and work from home. Vlogs may also be affected by monetization of he videos on YouTube. Despite these potential limitations, vlogs provide a unique form of observational data into a developer's natural environment. Compared to observations and field studies, vlogs can be studied at a much lower cost, and provide access to participants beyond physical boundaries and outside work hours. 

Additionally, it requires a level of comfort to be in-front of camera and capture certain parts of their day. In part, the goal of this research, specifically \ref{sec:result-rq2}, was to identify if there is an overlap in activities between those who the SE community refers to as an ``average'' developer (an LSC developer) versus developers that identify themselves beyond that (in this case software vloggers). Vlogs are a unique dataset that captures this diversity. Through this and future studies, we aim to understand how SE researchers and the community can broaden their perspective of who software developers are and what they do. 

The activity categories identified through negotiated agreement across three researchers are aggregated from ``A day in the life of a developer'' vlogs on \yt. These activities depend on many factors such as the job roles, location, and developers' individual experiences. Despite having a very diverse population of developers, our reported findings may not generalize to other types of videos by developers on other types of videos on \yt like tutorials or other platforms like TikTok. 

The \emph{survey} may have been influenced by self-selection bias~\cite{bethlehem2010selection}. When sharing preferences about what activities to capture in \pretendvlog, individual developers may have different priorities and feel more strongly about the value of certain activities than the average population of developers. Furthermore, the survey was conducted in a large software company and may not generalize to other companies or open source. What developers at LSC report they do are limited by response biases to some extent, influenced by the desire to present socially acceptable information. These are limitations of surveys in general. To reduce primacy bias we randomized items when possible. To reduce social desirability bias, all survey responses were anonymous and unlinked in order to give respondents confidence to answer freely--or to not answer at all (no questions were required).
To facilitate replication of this work, we provide \emph{supplemental material}.

\section{Related Work}

\paragraph{YouTube and work.} The concepts of self-presentation and identity on YouTube have been studied for several types of workers, for example dentists~\cite{knosel2011youtube}, nurses~\cite{kelly2012image}, or Uber drivers~\cite{chan2019becoming}. YouTube videos have also been investigated in the context of learning, for example for student nurses~\cite{clifton2011can}. In general, online videos have been helpful for learning new technical content. For example, technical content being available as interactive lessons through MOOCs have been helpful with teaching technical content in large course settings~\cite{christensen2013mooc,hadavand2018can}. In the context of work, funny videos have been found to have a positive influence on work-related well-being~\cite{janicke2019finding}. A common self-presentation technique on YouTube and other social media is \emph{micro-celebrity}~\cite{marwick2011tweet} in which ``people view themselves as a public persona to be consumed by others, use strategic intimacy to appeal to followers, and regard their audience as fans''~\cite{marwick2015you}.

\paragraph{Developers on social media.} The landscape of software engineering has been changed dramatically by social media~\cite{begel2010social,storey2014revolution,storey2010impact}. Past empirical studies have focused on the use of various social media for software development, for example, social coding on GitHub~\cite{dabbish2012social}; asking questions on social Q\&A sites such as StackOverflow~\cite{vasilescu2014social}; staying current~\cite{singer2014software}; following OSS projects~\cite{singer2014software,fang2020need}, and creating social movements~\cite{liu2017selfies} with Twitter; discussing security threats on Reddit~\cite{li2020developers}; programming mentorship communities on Twitch.tv~\cite{faas2018watch}, and sharing documentation and knowledge on YouTube~\cite{macleod2015code}. 
Many developers publish video tutorials about software development on YouTube. Research has recently started to extract relevant fragments~\cite{ponzanelli2016too} and source code~\cite{yadid2016extracting,khormi2020study} from tutorial videos to support software developers.

\paragraph{How developer spend their time.}
Several studies have investigated developers' workdays, daily activities, and work processes to gain insights about software development. Several studies relied on telemetry data that automatically records developers' interactions with a computer. For example, Aman et al.~\cite{amann:saner:2016} investigated 6,300 hours of work time from over 100 professional C\# developers in Visual Studio to identify how much is spend on activities such as code editing and execution or navigation. Astromskis et al.~\cite{astromskis2017patterns} analyzed 1,000 hours of activities of six developers within an IDE with respect to development activities, online help, and task switching. A limitation of telemetry-based studies is that they cannot capture time away from the computer. Several studies relied on self-reported data, typically collected through surveys or online diaries. For example, Ford et al.~\cite{ford2017characterizing} identified various personas and working styles of developers based on self-reported time spent on knowledge worker activities in a survey. Meyer et al.~\cite{meyer2019today} characterized productive workdays of developers based of self-reported time spent in a survey. Conceptually the day-in-the-life vlogs are a different form of diary. While existing studies bring out important aspects of workdays, our study of vlogs reveals ``who developers are'' beyond just nine-to-five at work. Our study captures unique aspects such as hobbies and side project, social activities, health, workout, and lifestyle of developers.

\paragraph{Gray literature.} Unlike published, formal literature in academic journals and conferences papers, gray literature is considered to be blog posts, videos, and white papers. Special literature review techniques have emerged that include gray literature~\cite{garousi2019guidelines}. Often these reviews are based on industrial literature like for example this review on the pains and gains of micro-services~\cite{soldani2018pains}. The day-in-the-life vlogs analyzed in this paper can be considered as a form of gray literature, which to the best of our knowledge has not yet been studied in software engineering.

\section{Conclusion \& Future Work}
Our study presents an empirical analysis of what developers who vlog share about their professional and personal life to the world, and what developers who don't vlog anticipate sharing about their full-life experience publicly.
We find that both groups of developers provide insights into true experiences despite differences in how much technical depth (e.g., coding, problem solving, etc.) and social depth (e.g., lifestyle, family time, mental health, etc.) they present. Together, they present a richer and more complete description of `what software developers do' in a day.
We envision future research in this area deepening our understanding of developer identity presentation and how vlogs can be a vehicle for personal self reflection and naive realism.

\section*{Acknowledgements}
\balance
We thank the developers who shared their stories on YouTube and the respondents to our survey. Souti Chattopadhyay worked on this project as a research intern in the Software Analysis and Intelligence Group at Microsoft Research. This work was partially supported by the National Science Foundation under Grant Number 2008089.

\bibliographystyle{ACM-Reference-Format}
\bibliography{main.bib}

\end{document}

%% file: lookup_table.tex
\lookupPut{M1_01}{1}
\lookupPut{M1_02}{2}
\lookupPut{M1_03}{3}
\lookupPut{M1_04}{4}
\lookupPut{M1_05}{5}
\lookupPut{M1_06}{6}
\lookupPut{M1_07}{7}
\lookupPut{M1_08}{8}
\lookupPut{M1_09}{9}
\lookupPut{M1_10}{10}
\lookupPut{M1_11}{11}
\lookupPut{M1_13}{12}
\lookupPut{M1_14}{13}
\lookupPut{M1_15}{14}
\lookupPut{M1_16}{15}
\lookupPut{M1_17}{16}
\lookupPut{M1_19}{17}
\lookupPut{M1_20}{18}
\lookupPut{M1_21}{19}
\lookupPut{M1_22}{20}
\lookupPut{M1_23}{21}
\lookupPut{M1_24}{22}
\lookupPut{M1_25}{23}
\lookupPut{M1_26}{24}
\lookupPut{M1_27}{25}
\lookupPut{M1_28}{26}
\lookupPut{M1_29}{27}
\lookupPut{M1_30}{28}
\lookupPut{M1_31}{29}
\lookupPut{M1_33}{30}
\lookupPut{M1_34}{31}
\lookupPut{M1_37}{32}
\lookupPut{M1_38}{33}
\lookupPut{M1_39}{34}
\lookupPut{M1_40}{35}
\lookupPut{M1_42}{36}
\lookupPut{M1_46}{37}
\lookupPut{M1_47}{38}
\lookupPut{M1_48}{39}
\lookupPut{M1_49}{40}
\lookupPut{M1_50}{41}
\lookupPut{M1_51}{42}
\lookupPut{M1_52}{43}
\lookupPut{M1_54}{44}
\lookupPut{M1_55}{45}
\lookupPut{M1_56}{46}
\lookupPut{M1_57}{47}
\lookupPut{M1_58}{48}
\lookupPut{M1_60}{49}
\lookupPut{M1_61}{50}
\lookupPut{M1_63}{51}
\lookupPut{M1_64}{52}
\lookupPut{M1_65}{53}
\lookupPut{M1_66}{54}
\lookupPut{M1_67}{55}
\lookupPut{M1_68}{56}
\lookupPut{M1_69}{57}
\lookupPut{M1_70}{58}
\lookupPut{M1_71}{59}
\lookupPut{M1_72}{60}
\lookupPut{M1_74}{61}
\lookupPut{M1_75}{62}
\lookupPut{M1_78}{63}
\lookupPut{M1_82}{64}
\lookupPut{M1_84}{65}
\lookupPut{M1_85}{66}
\lookupPut{M1_86}{67}
\lookupPut{M1_87}{68}
\lookupPut{M1_88}{69}
\lookupPut{M1_91}{70}
\lookupPut{M1_92}{71}
\lookupPut{M1_93}{72}
\lookupPut{M2_94}{73}
\lookupPut{M2_95}{74}
\lookupPut{M2_96}{75}
\lookupPut{M2_98}{76}
\lookupPut{M2_99}{77}
\lookupPut{M2_100}{78}
\lookupPut{M2_101}{79}
\lookupPut{M2_102}{80}
\lookupPut{M2_103}{81}
\lookupPut{M2_104}{82}
\lookupPut{M2_105}{83}
\lookupPut{M2_106}{84}
\lookupPut{M2_107}{85}
\lookupPut{M2_110}{86}
\lookupPut{M2_111}{87}
\lookupPut{M2_112}{88}
\lookupPut{M2_113}{89}
\lookupPut{M2_114}{90}
\lookupPut{M2_115}{91}
\lookupPut{M2_118}{92}
\lookupPut{M2_119}{93}
\lookupPut{M2_120}{94}
\lookupPut{M2_121}{95}
\lookupPut{M2_122}{96}
\lookupPut{M2_123}{97}
\lookupPut{M2_124}{98}
\lookupPut{M2_125}{99}
\lookupPut{M2_126}{100}
\lookupPut{M2_127}{101}
\lookupPut{M2_128}{102}
\lookupPut{M2_129}{103}
\lookupPut{M2_130}{104}
\lookupPut{M2_131}{105}
\lookupPut{M2_132}{106}
\lookupPut{M2_133}{107}
\lookupPut{M2_134}{108}
\lookupPut{M3_135}{109}
\lookupPut{M3_136}{110}
\lookupPut{M3_137}{111}
\lookupPut{M3_138}{112}
\lookupPut{M3_139}{113}
\lookupPut{M3_141}{114}
\lookupPut{M3_142}{115}
\lookupPut{M4_144}{116}
\lookupPut{M4_145}{117}
\lookupPut{M4_146}{118}
\lookupPut{M5_149}{119}
\lookupPut{M5_150}{120}
\lookupPut{M5_152}{121}
\lookupPut{M5_153}{122}
\lookupPut{MS_16}{123}
\lookupPut{MS_17}{124}
\lookupPut{MS_18}{125}
\lookupPut{MS_19}{126}
\lookupPut{MS_2}{127}
\lookupPut{MS_7}{128}
\lookupPut{MS_8}{129}
\lookupPut{MS_9}{130}

\lookupPut{M1}{1}
\lookupPut{M2}{2}
\lookupPut{M3}{3}
\lookupPut{M4}{4}
\lookupPut{M5}{5}
\lookupPut{M6}{6}
\lookupPut{M7}{7}
\lookupPut{M8}{8}
\lookupPut{M9}{9}
\lookupPut{M10}{10}
\lookupPut{M11}{11}
\lookupPut{M13}{12}
\lookupPut{M14}{13}
\lookupPut{M15}{14}
\lookupPut{M16}{15}
\lookupPut{M17}{16}
\lookupPut{M19}{17}
\lookupPut{M20}{18}
\lookupPut{M21}{19}
\lookupPut{M22}{20}
\lookupPut{M23}{21}
\lookupPut{M24}{22}
\lookupPut{M25}{23}
\lookupPut{M26}{24}
\lookupPut{M27}{25}
\lookupPut{M28}{26}
\lookupPut{M29}{27}
\lookupPut{M30}{28}
\lookupPut{M31}{29}
\lookupPut{M33}{30}
\lookupPut{M34}{31}
\lookupPut{M37}{32}
\lookupPut{M38}{33}
\lookupPut{M39}{34}
\lookupPut{M40}{35}
\lookupPut{M42}{36}
\lookupPut{M46}{37}
\lookupPut{M47}{38}
\lookupPut{M48}{39}
\lookupPut{M49}{40}
\lookupPut{M50}{41}
\lookupPut{M51}{42}
\lookupPut{M52}{43}
\lookupPut{M54}{44}
\lookupPut{M55}{45}
\lookupPut{M56}{46}
\lookupPut{M57}{47}
\lookupPut{M58}{48}
\lookupPut{M60}{49}
\lookupPut{M61}{50}
\lookupPut{M63}{51}
\lookupPut{M64}{52}
\lookupPut{M65}{53}
\lookupPut{M66}{54}
\lookupPut{M67}{55}
\lookupPut{M68}{56}
\lookupPut{M69}{57}
\lookupPut{M70}{58}
\lookupPut{M71}{59}
\lookupPut{M72}{60}
\lookupPut{M74}{61}
\lookupPut{M75}{62}
\lookupPut{M78}{63}
\lookupPut{M82}{64}
\lookupPut{M84}{65}
\lookupPut{M85}{66}
\lookupPut{M86}{67}
\lookupPut{M87}{68}
\lookupPut{M88}{69}
\lookupPut{M91}{70}
\lookupPut{M92}{71}
\lookupPut{M93}{72}
\lookupPut{M94}{73}
\lookupPut{M95}{74}
\lookupPut{M96}{75}
\lookupPut{M98}{76}
\lookupPut{M99}{77}
\lookupPut{M100}{78}
\lookupPut{M101}{79}
\lookupPut{M102}{80}
\lookupPut{M103}{81}
\lookupPut{M104}{82}
\lookupPut{M105}{83}
\lookupPut{M106}{84}
\lookupPut{M107}{85}
\lookupPut{M110}{86}
\lookupPut{M111}{87}
\lookupPut{M112}{88}
\lookupPut{M113}{89}
\lookupPut{M114}{90}
\lookupPut{M115}{91}
\lookupPut{M118}{92}
\lookupPut{M119}{93}
\lookupPut{M120}{94}
\lookupPut{M121}{95}
\lookupPut{M122}{96}
\lookupPut{M123}{97}
\lookupPut{M124}{98}
\lookupPut{M125}{99}
\lookupPut{M126}{100}
\lookupPut{M127}{101}
\lookupPut{M128}{102}
\lookupPut{M129}{103}
\lookupPut{M130}{104}
\lookupPut{M131}{105}
\lookupPut{M132}{106}
\lookupPut{M133}{107}
\lookupPut{M134}{108}
\lookupPut{M135}{109}
\lookupPut{M136}{110}
\lookupPut{M137}{111}
\lookupPut{M138}{112}
\lookupPut{M139}{113}
\lookupPut{M141}{114}
\lookupPut{M142}{115}
\lookupPut{M144}{116}
\lookupPut{M145}{117}
\lookupPut{M146}{118}
\lookupPut{M149}{119}
\lookupPut{M150}{120}
\lookupPut{M152}{121}
\lookupPut{M153}{122}

\lookupPut{MS16}{123}
\lookupPut{MS17}{124}
\lookupPut{MS18}{125}
\lookupPut{MS19}{126}
\lookupPut{MS2}{127}
\lookupPut{MS7}{128}
\lookupPut{MS8}{129}
\lookupPut{MS9}{130}

%% file: utils.tex
\usepackage{tabularx}
\usepackage{amsmath}
\usepackage{amssymb}
\usepackage{amsfonts}
\usepackage{algpseudocode,algorithm,algorithmicx} 
\usepackage{graphicx}
\usepackage{textcomp}
\usepackage[ampersand]{easylist}
\usepackage{multicol}
\usepackage{multirow}

\usepackage{threeparttable}
\usepackage{tabularx}

\usepackage{rotating}
\usepackage[roman]{parnotes}
\usepackage[many]{tcolorbox}
\usepackage{soul}
\usepackage{xcolor}
\usepackage{comment}
\usepackage{subcaption}
\usepackage{booktabs}
\usepackage{colortbl}
\usepackage{verbatim}

\usepackage{xspace}
\xspaceaddexceptions{]\}}

\usepackage{hyperref}
\usepackage{array}
\usepackage[inline]{enumitem}
\usepackage[export]{adjustbox}
\usepackage{longfbox}
\usepackage{commath}
\usepackage[normalem]{ulem}
\usepackage{mdframed,lipsum}
\usepackage{ifdraft}
\usepackage{balance}

\usepackage{makecell}

\usepackage{wrapfig}

\newif\ifdraft
\drafttrue %

%% file: aliases.tex
\usepackage{xcolor} %
\definecolor{darkgreen}{rgb}{0.05,0.5,0.05}

\usepackage{framed}
\usepackage[strict]{changepage}

\definecolor{PAblue}{RGB}{0,122,204}%

\newcommand{\quo}[2]{ 
	\vspace{-0.15cm}
	\def\FrameCommand{%
		\hspace{0pt}%
		{\color{PAblue}\vrule width 2pt}%
		{\color{white}\vrule width 2pt}%
		\colorbox{white}
	}%
	\MakeFramed{\advance\hsize-\width\FrameRestore}%
	\noindent\hspace{-4.55pt}%
	\begin{adjustwidth}{}{0pt}
		\vspace{-2pt}%
		``\emph{#1}'' ({#2})
		\vspace{-3pt}
	\end{adjustwidth}\endMakeFramed%
	\vspace{-0.15cm}
}

\newcommand{\inlinequote}[1]{``\emph{#1}''}

\newcommand{\myicon}[1]{\begin{tabular}{l}\includegraphics[width=0.05\linewidth]{#1}\end{tabular}}

\newcommand{\myiconinline}[1]{\setlength\intextsep{-3pt}\setlength\columnsep{0pt}\begin{wrapfigure}{l}{0.1\columnwidth}\includegraphics[width=0.1\columnwidth]{#1}\end{wrapfigure}}

\definecolor{boxcolor}{RGB}{238, 223, 204} %
\DeclareRobustCommand{\mybox}[2][gray!20]{%
\begin{tcolorbox}[   %
        breakable,
        left=0pt,
        right=0pt,
        top=0pt,
        bottom=0pt,
        colback=#1,
        colframe=black,
        width=\dimexpr\columnwidth\relax, 
        enlarge left by=0mm,
        boxsep=5pt,
        outer arc=4pt,
        boxrule=.5mm
        ]
        #2
\end{tcolorbox}
}

\newcommand{\RQA}{How do software vloggers present their daily routine in vlogs?\xspace}
\newcommand{\RQB}{How do industry developers self report their time spent in similar activities?\xspace}
\newcommand{\RQC}{What would industry developers find valuable to be presented about their daily life?\xspace}

\newcommand{\yt}{YouTube\xspace}

\newcommand{\MSdevs}{LSC developers\xspace}
\newcommand{\vlogdevs}{software vloggers\xspace}
\newcommand{\pretendvlog}{mock-up vlog\xspace}
\newcommand{\pretendvlogs}{mock-up vlogs\xspace}
\newcommand{\bigcompany}{large software company\xspace}
\newcommand{\wellnessactivities}{wellness activities\xspace}

%% file: tables/vlog_activities.tex
\begin{table}
\newcommand{\myactivity}[2]{\scriptsize{\color{blue}$\rightarrow$ \textbf{#1}:\ #2}}
\renewcommand{\myicon}[1]{\begin{minipage}[t]{0.1\columnwidth}\vspace{0pt}\includegraphics[width=\columnwidth]{#1}\end{minipage}}

\caption{The activity categories from the analysis pf the 130 vlogs. The activities of each category used in the survey are indicated in {\color{blue}$\rightarrow$ blue text}.}
\label{tab:vlog-activities}
\vspace{-0.5\baselineskip}

\begin{tabular}{@{}p{0.1\columnwidth}@{}p{0.9\columnwidth}@{}}
\toprule

\myicon{icons/noun_hacker_2767862.png} & \small \textsc{\textbf{Code Sessions}} \hfill  \textsc{(63 videos)}  \newline Working on code, and related activities like debugging, testing, and checking code into repository. Outside the editor, consulting external sources to problem-solve and reviewing code from juniors.

\vspace{2pt}
\myactivity{Coding}{Coding and reviewing (your or others') code}
\myactivity{Peripheral coding tasks}{Code related activities (like testing, refactoring, debugging, documentation etc.)}
\myactivity{Problem solving}{Investigating and designing a solution to a problem}

\\[-3pt]
\myicon{icons/noun_Work_3497804.png} & \small \textsc{\textbf{Unspecified Work}} \hfill \textsc{(54 videos)}\newline
Parts of video capturing developers at work without specifically disclosing the activities or codes, tools etc. Typically a time-lapse recording of the workspace.

\\[-3pt]
\myicon{icons/noun_administrative_506650.png} & \small \textsc{\textbf{Administrative Tasks}} \hfill \textsc{(19 videos)} \newline %
Preparing for and conducting interviews, mentoring interns and juniors, meeting and discussing with designers and architects. For freelance and remote developers, logging work time and handling clients.

\vspace{2pt}
\myactivity{Administrative work}{Doing administrative tasks like scheduling and organizing events and meetings}
\myactivity{Mentoring}{Mentoring and helping others around you}

\\[-3pt]
\myicon{icons/noun_organize_1519532.png} & \small \textsc{\textbf{Managing work/Planning the day}} \hfill \textsc{(34 videos)}\newline 
Organizing their workday on paper or apps, working with project management software like JIRA.

\vspace{2pt}
\myactivity{Organize}{Finding and using apps to increase productivity, make collaboration easier, organizing your day etc.}

\\[-3pt]
\myicon{icons/noun_Meeting_1603366.png} & \small \textsc{\textbf{Meetings}} \hfill \textsc{(65 videos)}\newline	
Attending and leading various meetings, presenting demonstrations in some meetings.

\vspace{2pt}
\myactivity{Meetings}{Having meetings (planned or unplanned) and hallway chats}

\\[-3pt]
\myicon{icons/noun_collaboration_3215610.png} & \small \textsc{\textbf{Collaboration}}\hfill \textsc{(41 videos)}\newline	
Working with immediate team members, designers, marketing team etc. Communicating with co-workers either in-person or through emails and messages.

\vspace{2pt}
\myactivity{Networking}{Networking with people outside of circle, building your personal brand}
\myactivity{Messaging}{Talking to colleagues through video calls, emails and messages}

\\[-3pt]
\myicon{icons/noun_Learning_3475260.png} & \small \textsc{\textbf{Learning}} \hfill \textsc{(12 videos)}\newline	
Discussing techniques to learn skills, working on online courses on Udemy, CodeAcademy etc.

\\[-3pt]
\myicon{icons/noun_creative_3049475.png} & \small \textsc{\textbf{Hobbies and Side Projects}} \hfill \textsc{(35 videos)}\newline	
Working on various technical projects (like building games and applications) and non-technical projects (like running a bakery, photography) outside of work hours, volunteering and teaching at welfare organizations.

\vspace{2pt}
\myactivity{Personal projects}{Working on your personal projects (game/app development, home improvement, music production, anything you are passionate about)}

\\[-3pt]
\myicon{icons/noun_Society_2830000.png} & \small \textsc{\textbf{Social Activities}} \hfill \textsc{(37 videos)}\newline
Spending time with family and kids at home, or with friends at social events or recreational sports.

\vspace{2pt}
\myactivity{Social}{With friends and family (game night, social hours, trips etc.)}

\\[-3pt]
\myicon{icons/noun_lifestyle_2514128.png} & \small \textsc{\textbf{Health, Workout, Lifestyle}} \hfill \textsc{(100 videos)}\newline
Working out or engaging in physical training, maintaining healthy and abundant food intakes, taking short breaks or light community games. Also practicing yoga, different mindfulness practices, and doing regular tasks like groceries and laundry.	

\vspace{2pt}
\myactivity{Lifestyle}{Lifestyle related activities (browsing through gadgets and technology, spending on home improvement and accessories, planning and travelling)}
\myactivity{Health}{Health related activities (workout, meditation)}
\myactivity{Breaks}{Taking breaks in between work} \\

\bottomrule

\end{tabular}
\end{table}

%% file: main.bbl

\begin{thebibliography}{50}

%%% the Web address.

\ifx \showCODEN    \undefined \def \showCODEN     #1{\unskip}     \fi
\ifx \showDOI      \undefined \def \showDOI       #1{#1}\fi
\ifx \showISBNx    \undefined \def \showISBNx     #1{\unskip}     \fi
\ifx \showISBNxiii \undefined \def \showISBNxiii  #1{\unskip}     \fi
\ifx \showISSN     \undefined \def \showISSN      #1{\unskip}     \fi
\ifx \showLCCN     \undefined \def \showLCCN      #1{\unskip}     \fi
\ifx \shownote     \undefined \def \shownote      #1{#1}          \fi
\ifx \showarticletitle \undefined \def \showarticletitle #1{#1}   \fi
\ifx \showURL      \undefined \def \showURL       {\relax}        \fi
% invisible to TeX
\providecommand\bibfield[2]{#2}
\providecommand\bibinfo[2]{#2}
\providecommand\natexlab[1]{#1}
\providecommand\showeprint[2][]{arXiv:#2}

\bibitem[\protect\citeauthoryear{??}{cod}{2021}]%
        {codeacademy}
 \bibinfo{year}{2021}\natexlab{}.
\newblock \bibinfo{booktitle}{\emph{Codeacademy. Learn data science, web
  development, programming, and more.}}
\newblock
\urldef\tempurl%
\url{https://www.codecademy.com/}
\showURL{%
\tempurl}


\bibitem[\protect\citeauthoryear{??}{ins}{2021}]%
        {instagram}
 \bibinfo{year}{2021}\natexlab{}.
\newblock \bibinfo{booktitle}{\emph{Instagram}}.
\newblock
\urldef\tempurl%
\url{https://www.instagram.com/}
\showURL{%
\tempurl}


\bibitem[\protect\citeauthoryear{??}{not}{2021}]%
        {notion}
 \bibinfo{year}{2021}\natexlab{}.
\newblock \bibinfo{booktitle}{\emph{Notion. All-in-one workspace. One tool for
  your whole team. Write, plan, and get organized.}}
\newblock
\urldef\tempurl%
\url{https://www.notion.so/}
\showURL{%
\tempurl}


\bibitem[\protect\citeauthoryear{??}{red}{2021}]%
        {redux}
 \bibinfo{year}{2021}\natexlab{}.
\newblock \bibinfo{booktitle}{\emph{Redux. A Predictable State Container for JS
  Apps}}.
\newblock
\urldef\tempurl%
\url{https://redux.js.org/}
\showURL{%
\tempurl}


\bibitem[\protect\citeauthoryear{??}{ste}{2021}]%
        {steam}
 \bibinfo{year}{2021}\natexlab{}.
\newblock \bibinfo{booktitle}{\emph{Steam. Steam is the ultimate destination
  for playing, discussing, and creating games.}}
\newblock
\urldef\tempurl%
\url{https://store.steampowered.com/about/}
\showURL{%
\tempurl}


\bibitem[\protect\citeauthoryear{??}{swi}{2021}]%
        {swiftui}
 \bibinfo{year}{2021}\natexlab{}.
\newblock \bibinfo{booktitle}{\emph{SwiftUI. Better Apps. Less Code.}}
\newblock
\urldef\tempurl%
\url{https://developer.apple.com/xcode/swiftui/}
\showURL{%
\tempurl}


\bibitem[\protect\citeauthoryear{??}{ude}{2021}]%
        {udemy}
 \bibinfo{year}{2021}\natexlab{}.
\newblock \bibinfo{booktitle}{\emph{Udemy. Improving Lives Through Learning}}.
\newblock
\urldef\tempurl%
\url{https://about.udemy.com/?locale=en-us}
\showURL{%
\tempurl}


\bibitem[\protect\citeauthoryear{??}{uni}{2021}]%
        {unity}
 \bibinfo{year}{2021}\natexlab{}.
\newblock \bibinfo{booktitle}{\emph{Unity. For all the creators. Break the
  barriers of reality -- bring new ideas to life with Unity.}}
\newblock
\urldef\tempurl%
\url{https://unity.com/}
\showURL{%
\tempurl}


\bibitem[\protect\citeauthoryear{{Amann}, {Proksch}, {Nadi}, and
  {Mezini}}{{Amann} et~al\mbox{.}}{2016}]%
        {amann:saner:2016}
\bibfield{author}{\bibinfo{person}{S. {Amann}}, \bibinfo{person}{S. {Proksch}},
  \bibinfo{person}{S. {Nadi}}, {and} \bibinfo{person}{M. {Mezini}}.}
  \bibinfo{year}{2016}\natexlab{}.
\newblock \showarticletitle{A Study of Visual Studio Usage in Practice}. In
  \bibinfo{booktitle}{\emph{2016 IEEE 23rd International Conference on Software
  Analysis, Evolution, and Reengineering (SANER)}}, Vol.~\bibinfo{volume}{1}.
  \bibinfo{pages}{124--134}.
\newblock
\urldef\tempurl%
\url{https://doi.org/10.1109/SANER.2016.39}
\showDOI{\tempurl}


\bibitem[\protect\citeauthoryear{Aran, Biel, and Gatica-Perez}{Aran
  et~al\mbox{.}}{2013}]%
        {aran2013broadcasting}
\bibfield{author}{\bibinfo{person}{Oya Aran}, \bibinfo{person}{Joan-Isaac
  Biel}, {and} \bibinfo{person}{Daniel Gatica-Perez}.}
  \bibinfo{year}{2013}\natexlab{}.
\newblock \showarticletitle{Broadcasting oneself: Visual discovery of vlogging
  styles}.
\newblock \bibinfo{journal}{\emph{IEEE Transactions on multimedia}}
  \bibinfo{volume}{16}, \bibinfo{number}{1} (\bibinfo{year}{2013}),
  \bibinfo{pages}{201--215}.
\newblock
\urldef\tempurl%
\url{https://doi.org/10.1109/TMM.2013.2284893}
\showDOI{\tempurl}


\bibitem[\protect\citeauthoryear{Astromskis, Bavota, Janes, Russo, and
  Di~Penta}{Astromskis et~al\mbox{.}}{2017}]%
        {astromskis2017patterns}
\bibfield{author}{\bibinfo{person}{Saulius Astromskis},
  \bibinfo{person}{Gabriele Bavota}, \bibinfo{person}{Andrea Janes},
  \bibinfo{person}{Barbara Russo}, {and} \bibinfo{person}{Massimiliano
  Di~Penta}.} \bibinfo{year}{2017}\natexlab{}.
\newblock \showarticletitle{Patterns of developers behaviour: A 1000-hour
  industrial study}.
\newblock \bibinfo{journal}{\emph{Journal of Systems and Software}}
  \bibinfo{volume}{132} (\bibinfo{year}{2017}), \bibinfo{pages}{85--97}.
\newblock
\urldef\tempurl%
\url{https://doi.org/10.1016/j.jss.2017.06.072}
\showURL{%
\tempurl}


\bibitem[\protect\citeauthoryear{Atlassian}{Atlassian}{2021}]%
        {JIRA}
\bibfield{author}{\bibinfo{person}{Atlassian}.}
  \bibinfo{year}{2021}\natexlab{}.
\newblock \bibinfo{booktitle}{\emph{Issue and Project Tracking Software}}.
\newblock
\urldef\tempurl%
\url{https://www.atlassian.com/software/jira}
\showURL{%
\tempurl}


\bibitem[\protect\citeauthoryear{Baumeister and Hutton}{Baumeister and
  Hutton}{1987}]%
        {baumeister1987selfpresentation}
\bibfield{author}{\bibinfo{person}{Roy~F Baumeister} {and}
  \bibinfo{person}{Debra~G Hutton}.} \bibinfo{year}{1987}\natexlab{}.
\newblock \showarticletitle{Self-presentation theory: Self-construction and
  audience pleasing}.
\newblock In \bibinfo{booktitle}{\emph{Theories of group behavior}}.
  \bibinfo{publisher}{Springer}, \bibinfo{pages}{71--87}.
\newblock
\urldef\tempurl%
\url{https://doi.org/10.1007/978-1-4612-4634-3_4}
\showDOI{\tempurl}


\bibitem[\protect\citeauthoryear{Begel, DeLine, and Zimmermann}{Begel
  et~al\mbox{.}}{2010}]%
        {begel2010social}
\bibfield{author}{\bibinfo{person}{Andrew Begel}, \bibinfo{person}{Robert
  DeLine}, {and} \bibinfo{person}{Thomas Zimmermann}.}
  \bibinfo{year}{2010}\natexlab{}.
\newblock \showarticletitle{Social media for software engineering}. In
  \bibinfo{booktitle}{\emph{Proceedings of the FSE/SDP workshop on Future of
  software engineering research}}. \bibinfo{pages}{33--38}.
\newblock
\urldef\tempurl%
\url{https://doi.org/10.1145/1882362.1882370}
\showDOI{\tempurl}


\bibitem[\protect\citeauthoryear{Bethlehem}{Bethlehem}{2010}]%
        {bethlehem2010selection}
\bibfield{author}{\bibinfo{person}{Jelke Bethlehem}.}
  \bibinfo{year}{2010}\natexlab{}.
\newblock \showarticletitle{Selection bias in web surveys}.
\newblock \bibinfo{journal}{\emph{International Statistical Review}}
  \bibinfo{volume}{78}, \bibinfo{number}{2} (\bibinfo{year}{2010}),
  \bibinfo{pages}{161--188}.
\newblock
\urldef\tempurl%
\url{https://doi.org/10.1111/j.1751-5823.2010.00112.x}
\showURL{%
\tempurl}


\bibitem[\protect\citeauthoryear{Chan}{Chan}{2019}]%
        {chan2019becoming}
\bibfield{author}{\bibinfo{person}{Ngai~Keung Chan}.}
  \bibinfo{year}{2019}\natexlab{}.
\newblock \showarticletitle{“Becoming an expert in driving for Uber”: Uber
  driver/bloggers’ performance of expertise and self-presentation on
  YouTube}.
\newblock \bibinfo{journal}{\emph{New Media \& Society}} \bibinfo{volume}{21},
  \bibinfo{number}{9} (\bibinfo{year}{2019}), \bibinfo{pages}{2048--2067}.
\newblock
\urldef\tempurl%
\url{https://doi.org/10.1177/1461444819837736}
\showURL{%
\tempurl}


\bibitem[\protect\citeauthoryear{Christensen, Steinmetz, Alcorn, Bennett,
  Woods, and Emanuel}{Christensen et~al\mbox{.}}{2013}]%
        {christensen2013mooc}
\bibfield{author}{\bibinfo{person}{Gayle Christensen}, \bibinfo{person}{Andrew
  Steinmetz}, \bibinfo{person}{Brandon Alcorn}, \bibinfo{person}{Amy Bennett},
  \bibinfo{person}{Deirdre Woods}, {and} \bibinfo{person}{Ezekiel Emanuel}.}
  \bibinfo{year}{2013}\natexlab{}.
\newblock \showarticletitle{The MOOC phenomenon: who takes massive open online
  courses and why?}
\newblock \bibinfo{journal}{\emph{SSRN}} (\bibinfo{year}{2013}).
\newblock
\urldef\tempurl%
\url{https://doi.org/10.2139/ssrn.2350964}
\showDOI{\tempurl}


\bibitem[\protect\citeauthoryear{Clement}{Clement}{2019}]%
        {clement2019youtube}
\bibfield{author}{\bibinfo{person}{J. Clement}.}
  \bibinfo{year}{2019}\natexlab{}.
\newblock \bibinfo{title}{Average YouTube video length as of December 2018, by
  category}.
\newblock
\newblock
\newblock
\shownote{Retrieved Sept 20, 2020 from
  \url{https://www.statista.com/statistics/1026923/youtube-video-category-average-length/}.}


\bibitem[\protect\citeauthoryear{Clifton and Mann}{Clifton and Mann}{2011}]%
        {clifton2011can}
\bibfield{author}{\bibinfo{person}{Andrew Clifton} {and}
  \bibinfo{person}{Claire Mann}.} \bibinfo{year}{2011}\natexlab{}.
\newblock \showarticletitle{Can YouTube enhance student nurse learning?}
\newblock \bibinfo{journal}{\emph{Nurse education today}} \bibinfo{volume}{31},
  \bibinfo{number}{4} (\bibinfo{year}{2011}), \bibinfo{pages}{311--313}.
\newblock
\urldef\tempurl%
\url{https://doi.org/10.1016/j.nedt.2010.10.004}
\showURL{%
\tempurl}


\bibitem[\protect\citeauthoryear{Cummings, Huff, Mack, Womack, Reid, Ghoram,
  Gilbert, and Gosha}{Cummings et~al\mbox{.}}{2019}]%
        {cummings2019vlog}
\bibfield{author}{\bibinfo{person}{Robert Cummings}, \bibinfo{person}{Earl
  Huff}, \bibinfo{person}{Naja Mack}, \bibinfo{person}{Kevin Womack},
  \bibinfo{person}{Amber Reid}, \bibinfo{person}{Brandon Ghoram},
  \bibinfo{person}{Juan Gilbert}, {and} \bibinfo{person}{Kinnis Gosha}.}
  \bibinfo{year}{2019}\natexlab{}.
\newblock \showarticletitle{Vlog Commentary YouTube Influencers as Effective
  Advisors in College and Career Readiness for Minorities in Computing: An
  Exploratory Study}. In \bibinfo{booktitle}{\emph{2019 Research on Equity and
  Sustained Participation in Engineering, Computing, and Technology
  (RESPECT)}}. \bibinfo{pages}{1--8}.
\newblock
\urldef\tempurl%
\url{https://doi.org/10.1109/RESPECT46404.2019.8985961}
\showDOI{\tempurl}


\bibitem[\protect\citeauthoryear{Dabbish, Stuart, Tsay, and Herbsleb}{Dabbish
  et~al\mbox{.}}{2012}]%
        {dabbish2012social}
\bibfield{author}{\bibinfo{person}{Laura Dabbish}, \bibinfo{person}{Colleen
  Stuart}, \bibinfo{person}{Jason Tsay}, {and} \bibinfo{person}{Jim Herbsleb}.}
  \bibinfo{year}{2012}\natexlab{}.
\newblock \showarticletitle{Social coding in GitHub: transparency and
  collaboration in an open software repository}. In
  \bibinfo{booktitle}{\emph{Proceedings of the ACM 2012 conference on computer
  supported cooperative work}}. \bibinfo{pages}{1277--1286}.
\newblock
\urldef\tempurl%
\url{https://doi.org/10.1145/2145204.2145396}
\showURL{%
\tempurl}


\bibitem[\protect\citeauthoryear{Elizarova, Briselli, and Dowd}{Elizarova
  et~al\mbox{.}}{2017}]%
        {elizarova2017participatory}
\bibfield{author}{\bibinfo{person}{Olga Elizarova}, \bibinfo{person}{Jen
  Briselli}, {and} \bibinfo{person}{Kimberly Dowd}.}
  \bibinfo{year}{2017}\natexlab{}.
\newblock \bibinfo{title}{Participatory Design in Practice}.
\newblock
\newblock
\newblock
\shownote{Retrieved February 19, 2021 from
  \url{https://uxmag.com/articles/participatory-design-in-practice}.}


\bibitem[\protect\citeauthoryear{Faas, Dombrowski, Young, and Miller}{Faas
  et~al\mbox{.}}{2018}]%
        {faas2018watch}
\bibfield{author}{\bibinfo{person}{Travis Faas}, \bibinfo{person}{Lynn
  Dombrowski}, \bibinfo{person}{Alyson Young}, {and} \bibinfo{person}{Andrew~D
  Miller}.} \bibinfo{year}{2018}\natexlab{}.
\newblock \showarticletitle{Watch me code: Programming mentorship communities
  on Twitch.tv}.
\newblock \bibinfo{journal}{\emph{Proceedings of the ACM on Human-Computer
  Interaction}} \bibinfo{volume}{2}, \bibinfo{number}{CSCW}
  (\bibinfo{year}{2018}), \bibinfo{pages}{1--18}.
\newblock
\urldef\tempurl%
\url{https://doi.org/10.1145/3274319}
\showDOI{\tempurl}


\bibitem[\protect\citeauthoryear{Fang, Klug, Lamba, Herbsleb, and
  Vasilescu}{Fang et~al\mbox{.}}{2020}]%
        {fang2020need}
\bibfield{author}{\bibinfo{person}{Hongbo Fang}, \bibinfo{person}{Daniel Klug},
  \bibinfo{person}{Hemank Lamba}, \bibinfo{person}{James Herbsleb}, {and}
  \bibinfo{person}{Bogdan Vasilescu}.} \bibinfo{year}{2020}\natexlab{}.
\newblock \showarticletitle{Need for Tweet: How Open Source Developers Talk
  About Their GitHub Work on Twitter}. In \bibinfo{booktitle}{\emph{Proceedings
  of the 17th International Conference on Mining Software Repositories}}
  (Seoul, Republic of Korea) \emph{(\bibinfo{series}{MSR '20})}.
  \bibinfo{publisher}{Association for Computing Machinery},
  \bibinfo{address}{New York, NY, USA}, \bibinfo{pages}{322–326}.
\newblock
\showISBNx{9781450375177}
\urldef\tempurl%
\url{https://doi.org/10.1145/3379597.3387466}
\showDOI{\tempurl}


\bibitem[\protect\citeauthoryear{Ford, Zimmermann, Bird, and Nagappan}{Ford
  et~al\mbox{.}}{2017}]%
        {ford2017characterizing}
\bibfield{author}{\bibinfo{person}{Denae Ford}, \bibinfo{person}{Thomas
  Zimmermann}, \bibinfo{person}{Christian Bird}, {and}
  \bibinfo{person}{Nachiappan Nagappan}.} \bibinfo{year}{2017}\natexlab{}.
\newblock \showarticletitle{Characterizing Software Engineering Work with
  Personas Based on Knowledge Worker Actions}. In
  \bibinfo{booktitle}{\emph{Proceedings of the 11th ACM/IEEE International
  Symposium on Empirical Software Engineering and Measurement}} (Markham,
  Ontario, Canada) \emph{(\bibinfo{series}{ESEM '17})}.
  \bibinfo{publisher}{IEEE Press}, \bibinfo{pages}{394–403}.
\newblock
\showISBNx{9781509040391}
\urldef\tempurl%
\url{https://doi.org/10.1109/ESEM.2017.54}
\showDOI{\tempurl}


\bibitem[\protect\citeauthoryear{Garousi, Felderer, and
  M{\"a}ntyl{\"a}}{Garousi et~al\mbox{.}}{2019}]%
        {garousi2019guidelines}
\bibfield{author}{\bibinfo{person}{Vahid Garousi}, \bibinfo{person}{Michael
  Felderer}, {and} \bibinfo{person}{Mika~V M{\"a}ntyl{\"a}}.}
  \bibinfo{year}{2019}\natexlab{}.
\newblock \showarticletitle{Guidelines for including grey literature and
  conducting multivocal literature reviews in software engineering}.
\newblock \bibinfo{journal}{\emph{Information and Software Technology}}
  \bibinfo{volume}{106} (\bibinfo{year}{2019}), \bibinfo{pages}{101--121}.
\newblock
\urldef\tempurl%
\url{https://doi.org/10.1016/j.infsof.2018.09.006}
\showURL{%
\tempurl}


\bibitem[\protect\citeauthoryear{Hadavand, Gooding, and Leek}{Hadavand
  et~al\mbox{.}}{2018}]%
        {hadavand2018can}
\bibfield{author}{\bibinfo{person}{Aboozar Hadavand}, \bibinfo{person}{Ira
  Gooding}, {and} \bibinfo{person}{Jeffrey~T Leek}.}
  \bibinfo{year}{2018}\natexlab{}.
\newblock \showarticletitle{Can MOOC Programs Improve Student Employment
  Prospects?}
\newblock \bibinfo{journal}{\emph{SSRN}} (\bibinfo{year}{2018}).
\newblock
\urldef\tempurl%
\url{https://doi.org/10.2139/ssrn.3260695}
\showDOI{\tempurl}


\bibitem[\protect\citeauthoryear{Janicke-Bowles, Rieger, and
  Connor}{Janicke-Bowles et~al\mbox{.}}{2019}]%
        {janicke2019finding}
\bibfield{author}{\bibinfo{person}{Sophie~H Janicke-Bowles},
  \bibinfo{person}{Diana Rieger}, {and} \bibinfo{person}{Winston Connor}.}
  \bibinfo{year}{2019}\natexlab{}.
\newblock \showarticletitle{Finding meaning at work: The role of inspiring and
  funny YouTube videos on work-related well-being}.
\newblock \bibinfo{journal}{\emph{Journal of Happiness Studies}}
  \bibinfo{volume}{20}, \bibinfo{number}{2} (\bibinfo{year}{2019}),
  \bibinfo{pages}{619--640}.
\newblock
\urldef\tempurl%
\url{https://doi.org/10.1007/s10902-018-9959-1}
\showURL{%
\tempurl}


\bibitem[\protect\citeauthoryear{Jensen~Schau and Gilly}{Jensen~Schau and
  Gilly}{2003}]%
        {jensen2003selfpresentation}
\bibfield{author}{\bibinfo{person}{Hope Jensen~Schau} {and}
  \bibinfo{person}{Mary~C Gilly}.} \bibinfo{year}{2003}\natexlab{}.
\newblock \showarticletitle{We are what we post? Self-presentation in personal
  web space}.
\newblock \bibinfo{journal}{\emph{Journal of consumer research}}
  \bibinfo{volume}{30}, \bibinfo{number}{3} (\bibinfo{year}{2003}),
  \bibinfo{pages}{385--404}.
\newblock
\urldef\tempurl%
\url{https://doi.org/10.1086/378616}
\showDOI{\tempurl}


\bibitem[\protect\citeauthoryear{Kelly, Fealy, and Watson}{Kelly
  et~al\mbox{.}}{2012}]%
        {kelly2012image}
\bibfield{author}{\bibinfo{person}{Jacinta Kelly}, \bibinfo{person}{Gerard~M
  Fealy}, {and} \bibinfo{person}{Roger Watson}.}
  \bibinfo{year}{2012}\natexlab{}.
\newblock \showarticletitle{The image of you: constructing nursing identities
  in YouTube}.
\newblock \bibinfo{journal}{\emph{Journal of Advanced Nursing}}
  \bibinfo{volume}{68}, \bibinfo{number}{8} (\bibinfo{year}{2012}),
  \bibinfo{pages}{1804--1813}.
\newblock
\urldef\tempurl%
\url{https://doi.org/10.1111/j.1365-2648.2011.05872.x}
\showURL{%
\tempurl}


\bibitem[\protect\citeauthoryear{Khormi, Alahmadi, and Haiduc}{Khormi
  et~al\mbox{.}}{2020}]%
        {khormi2020study}
\bibfield{author}{\bibinfo{person}{Abdulkarim Khormi},
  \bibinfo{person}{Mohammad Alahmadi}, {and} \bibinfo{person}{Sonia Haiduc}.}
  \bibinfo{year}{2020}\natexlab{}.
\newblock \showarticletitle{A study on the accuracy of ocr engines for source
  code transcription from programming screencasts}. In
  \bibinfo{booktitle}{\emph{Proceedings of the 17th International Conference on
  Mining Software Repositories}}. \bibinfo{pages}{65--75}.
\newblock
\urldef\tempurl%
\url{https://doi.org/10.1145/3379597.3387468}
\showURL{%
\tempurl}


\bibitem[\protect\citeauthoryear{Kn{\"o}sel, Jung, and Bleckmann}{Kn{\"o}sel
  et~al\mbox{.}}{2011}]%
        {knosel2011youtube}
\bibfield{author}{\bibinfo{person}{Michael Kn{\"o}sel}, \bibinfo{person}{Klaus
  Jung}, {and} \bibinfo{person}{Annalen Bleckmann}.}
  \bibinfo{year}{2011}\natexlab{}.
\newblock \showarticletitle{YouTube, dentistry, and dental education}.
\newblock \bibinfo{journal}{\emph{Journal of dental education}}
  \bibinfo{volume}{75}, \bibinfo{number}{12} (\bibinfo{year}{2011}),
  \bibinfo{pages}{1558--1568}.
\newblock
\urldef\tempurl%
\url{https://doi.org/10.1002/j.0022-0337.2011.75.12.tb05215.x}
\showURL{%
\tempurl}


\bibitem[\protect\citeauthoryear{Li, Louie, Dabbish, and Hong}{Li
  et~al\mbox{.}}{2021}]%
        {li2020developers}
\bibfield{author}{\bibinfo{person}{Tianshi Li}, \bibinfo{person}{Elizabeth
  Louie}, \bibinfo{person}{Laura Dabbish}, {and} \bibinfo{person}{Jason~I.
  Hong}.} \bibinfo{year}{2021}\natexlab{}.
\newblock \showarticletitle{How Developers Talk About Personal Data and What It
  Means for User Privacy: A Case Study of a Developer Forum on Reddit}.
\newblock \bibinfo{journal}{\emph{Proc. ACM Hum.-Comput. Interact.}}
  \bibinfo{volume}{4}, \bibinfo{number}{CSCW3}, Article
  \bibinfo{articleno}{220} (\bibinfo{date}{Jan.} \bibinfo{year}{2021}),
  \bibinfo{numpages}{28}~pages.
\newblock
\urldef\tempurl%
\url{https://doi.org/10.1145/3432919}
\showDOI{\tempurl}


\bibitem[\protect\citeauthoryear{Liu, Ford, Parnin, and Dabbish}{Liu
  et~al\mbox{.}}{2017}]%
        {liu2017selfies}
\bibfield{author}{\bibinfo{person}{Fannie Liu}, \bibinfo{person}{Denae Ford},
  \bibinfo{person}{Chris Parnin}, {and} \bibinfo{person}{Laura Dabbish}.}
  \bibinfo{year}{2017}\natexlab{}.
\newblock \showarticletitle{Selfies as Social Movements: Influences on
  Participation and Perceived Impact on Stereotypes}.
\newblock \bibinfo{journal}{\emph{Proc. ACM Hum.-Comput. Interact.}}
  \bibinfo{volume}{1}, \bibinfo{number}{CSCW}, Article \bibinfo{articleno}{72}
  (\bibinfo{date}{Dec.} \bibinfo{year}{2017}), \bibinfo{numpages}{21}~pages.
\newblock
\urldef\tempurl%
\url{https://doi.org/10.1145/3134707}
\showDOI{\tempurl}


\bibitem[\protect\citeauthoryear{MacLeod, Storey, and Bergen}{MacLeod
  et~al\mbox{.}}{2015}]%
        {macleod2015code}
\bibfield{author}{\bibinfo{person}{Laura MacLeod},
  \bibinfo{person}{Margaret-Anne Storey}, {and} \bibinfo{person}{Andreas
  Bergen}.} \bibinfo{year}{2015}\natexlab{}.
\newblock \showarticletitle{Code, camera, action: How software developers
  document and share program knowledge using YouTube}. In
  \bibinfo{booktitle}{\emph{2015 IEEE 23rd International Conference on Program
  Comprehension}}. IEEE, \bibinfo{pages}{104--114}.
\newblock
\urldef\tempurl%
\url{https://doi.org/10.1109/ICPC.2015.19}
\showDOI{\tempurl}


\bibitem[\protect\citeauthoryear{Marwick}{Marwick}{2015}]%
        {marwick2015you}
\bibfield{author}{\bibinfo{person}{Alice Marwick}.}
  \bibinfo{year}{2015}\natexlab{}.
\newblock \showarticletitle{You may know me from YouTube}.
\newblock \bibinfo{journal}{\emph{A companion to celebrity}}
  \bibinfo{volume}{333} (\bibinfo{year}{2015}).
\newblock
\urldef\tempurl%
\url{https://doi.org/10.1002/9781118475089.ch18}
\showURL{%
\tempurl}


\bibitem[\protect\citeauthoryear{Marwick and Boyd}{Marwick and Boyd}{2011}]%
        {marwick2011tweet}
\bibfield{author}{\bibinfo{person}{Alice~E Marwick} {and}
  \bibinfo{person}{Danah Boyd}.} \bibinfo{year}{2011}\natexlab{}.
\newblock \showarticletitle{I tweet honestly, I tweet passionately: Twitter
  users, context collapse, and the imagined audience}.
\newblock \bibinfo{journal}{\emph{New media \& society}} \bibinfo{volume}{13},
  \bibinfo{number}{1} (\bibinfo{year}{2011}), \bibinfo{pages}{114--133}.
\newblock
\urldef\tempurl%
\url{https://doi.org/10.1177/1461444810365313}
\showURL{%
\tempurl}


\bibitem[\protect\citeauthoryear{Meyer, Barr, Bird, and Zimmermann}{Meyer
  et~al\mbox{.}}{2019}]%
        {meyer2019today}
\bibfield{author}{\bibinfo{person}{Andre Meyer}, \bibinfo{person}{Earl~T Barr},
  \bibinfo{person}{Christian Bird}, {and} \bibinfo{person}{Thomas Zimmermann}.}
  \bibinfo{year}{2019}\natexlab{}.
\newblock \showarticletitle{Today was a good day: The daily life of software
  developers}.
\newblock \bibinfo{journal}{\emph{IEEE Transactions on Software Engineering}}
  (\bibinfo{year}{2019}).
\newblock
\urldef\tempurl%
\url{https://doi.org/10.1109/TSE.2019.2904957}
\showDOI{\tempurl}


\bibitem[\protect\citeauthoryear{Ponzanelli, Bavota, Mocci, Di~Penta, Oliveto,
  Hasan, Russo, Haiduc, and Lanza}{Ponzanelli et~al\mbox{.}}{2016}]%
        {ponzanelli2016too}
\bibfield{author}{\bibinfo{person}{Luca Ponzanelli}, \bibinfo{person}{Gabriele
  Bavota}, \bibinfo{person}{Andrea Mocci}, \bibinfo{person}{Massimiliano
  Di~Penta}, \bibinfo{person}{Rocco Oliveto}, \bibinfo{person}{Mir Hasan},
  \bibinfo{person}{Barbara Russo}, \bibinfo{person}{Sonia Haiduc}, {and}
  \bibinfo{person}{Michele Lanza}.} \bibinfo{year}{2016}\natexlab{}.
\newblock \showarticletitle{Too long; didn't watch! extracting relevant
  fragments from software development video tutorials}. In
  \bibinfo{booktitle}{\emph{Proceedings of the 38th International Conference on
  Software Engineering}}. \bibinfo{pages}{261--272}.
\newblock
\urldef\tempurl%
\url{https://doi.org/10.1145/2884781.2884824}
\showURL{%
\tempurl}


\bibitem[\protect\citeauthoryear{Salda{\~n}a}{Salda{\~n}a}{2009}]%
        {Saldana2009}
\bibfield{author}{\bibinfo{person}{Johnny Salda{\~n}a}.}
  \bibinfo{year}{2009}\natexlab{}.
\newblock \bibinfo{booktitle}{\emph{The {{Coding Manual}} for {{Qualitative
  Researchers}}}}.
\newblock \bibinfo{publisher}{{SAGE Publications}}.
\newblock
\showISBNx{978-1-84787-548-8}


\bibitem[\protect\citeauthoryear{Singer, Figueira~Filho, and Storey}{Singer
  et~al\mbox{.}}{2014}]%
        {singer2014software}
\bibfield{author}{\bibinfo{person}{Leif Singer}, \bibinfo{person}{Fernando
  Figueira~Filho}, {and} \bibinfo{person}{Margaret-Anne Storey}.}
  \bibinfo{year}{2014}\natexlab{}.
\newblock \showarticletitle{Software engineering at the speed of light: how
  developers stay current using twitter}. In
  \bibinfo{booktitle}{\emph{Proceedings of the 36th International Conference on
  Software Engineering}}. \bibinfo{pages}{211--221}.
\newblock
\urldef\tempurl%
\url{https://doi.org/10.1145/2568225.2568305}
\showURL{%
\tempurl}


\bibitem[\protect\citeauthoryear{Smith, Loftin, Murphy-Hill, Bird, and
  Zimmermann}{Smith et~al\mbox{.}}{2013}]%
        {smith13}
\bibfield{author}{\bibinfo{person}{Edward Smith}, \bibinfo{person}{Robert
  Loftin}, \bibinfo{person}{Emerson Murphy-Hill}, \bibinfo{person}{Christian
  Bird}, {and} \bibinfo{person}{Thomas Zimmermann}.}
  \bibinfo{year}{2013}\natexlab{}.
\newblock \showarticletitle{Improving developer participation rates in
  surveys}. In \bibinfo{booktitle}{\emph{2013 6th International Workshop on
  Cooperative and Human Aspects of Software Engineering (CHASE)}}.
  \bibinfo{publisher}{IEEE}, \bibinfo{pages}{89--92}.
\newblock
\urldef\tempurl%
\url{https://doi.org/10.1109/CHASE.2013.6614738}
\showDOI{\tempurl}


\bibitem[\protect\citeauthoryear{Soldani, Tamburri, and Van Den~Heuvel}{Soldani
  et~al\mbox{.}}{2018}]%
        {soldani2018pains}
\bibfield{author}{\bibinfo{person}{Jacopo Soldani},
  \bibinfo{person}{Damian~Andrew Tamburri}, {and} \bibinfo{person}{Willem-Jan
  Van Den~Heuvel}.} \bibinfo{year}{2018}\natexlab{}.
\newblock \showarticletitle{The pains and gains of microservices: A Systematic
  grey literature review}.
\newblock \bibinfo{journal}{\emph{Journal of Systems and Software}}
  \bibinfo{volume}{146} (\bibinfo{year}{2018}), \bibinfo{pages}{215--232}.
\newblock
\urldef\tempurl%
\url{https://doi.org/10.1016/j.jss.2018.09.082}
\showURL{%
\tempurl}


\bibitem[\protect\citeauthoryear{Souti~Chattopadhyay}{Souti~Chattopadhyay}{2021}]%
        {supplementarymaterials}
\bibfield{author}{\bibinfo{person}{Denae~Ford Souti~Chattopadhyay,
  Thomas~Zimmermann}.} \bibinfo{year}{2021}\natexlab{}.
\newblock \bibinfo{title}{Supplemental Materials for Reel Life vs. Real Life:
  How Software Developers Share their Daily Life through Vlogs}.
\newblock
\newblock
\newblock
\shownote{Retrieved February 25, 2021 from
  \url{https://figshare.com/s/343f449568a0dc6ba5c4}.}


\bibitem[\protect\citeauthoryear{Storey, Singer, Cleary, Figueira~Filho, and
  Zagalsky}{Storey et~al\mbox{.}}{2014}]%
        {storey2014revolution}
\bibfield{author}{\bibinfo{person}{Margaret-Anne Storey}, \bibinfo{person}{Leif
  Singer}, \bibinfo{person}{Brendan Cleary}, \bibinfo{person}{Fernando
  Figueira~Filho}, {and} \bibinfo{person}{Alexey Zagalsky}.}
  \bibinfo{year}{2014}\natexlab{}.
\newblock \showarticletitle{The (R) Evolution of Social Media in Software
  Engineering}. In \bibinfo{booktitle}{\emph{Future of Software Engineering
  Proceedings}} (Hyderabad, India) \emph{(\bibinfo{series}{FOSE 2014})}.
  \bibinfo{publisher}{Association for Computing Machinery},
  \bibinfo{address}{New York, NY, USA}, \bibinfo{pages}{100–116}.
\newblock
\showISBNx{9781450328654}
\urldef\tempurl%
\url{https://doi.org/10.1145/2593882.2593887}
\showDOI{\tempurl}


\bibitem[\protect\citeauthoryear{Storey, Treude, van Deursen, and Cheng}{Storey
  et~al\mbox{.}}{2010}]%
        {storey2010impact}
\bibfield{author}{\bibinfo{person}{Margaret-Anne Storey},
  \bibinfo{person}{Christoph Treude}, \bibinfo{person}{Arie van Deursen}, {and}
  \bibinfo{person}{Li-Te Cheng}.} \bibinfo{year}{2010}\natexlab{}.
\newblock \showarticletitle{The impact of social media on software engineering
  practices and tools}. In \bibinfo{booktitle}{\emph{Proceedings of the FSE/SDP
  workshop on Future of software engineering research}}.
  \bibinfo{pages}{359--364}.
\newblock
\urldef\tempurl%
\url{https://doi.org/10.1145/1882362.1882435}
\showDOI{\tempurl}


\bibitem[\protect\citeauthoryear{Vasilescu, Serebrenik, Devanbu, and
  Filkov}{Vasilescu et~al\mbox{.}}{2014}]%
        {vasilescu2014social}
\bibfield{author}{\bibinfo{person}{Bogdan Vasilescu},
  \bibinfo{person}{Alexander Serebrenik}, \bibinfo{person}{Prem Devanbu}, {and}
  \bibinfo{person}{Vladimir Filkov}.} \bibinfo{year}{2014}\natexlab{}.
\newblock \showarticletitle{How social Q\&A sites are changing knowledge
  sharing in open source software communities}. In
  \bibinfo{booktitle}{\emph{Proceedings of the 17th ACM conference on Computer
  supported cooperative work \& social computing}}. \bibinfo{pages}{342--354}.
\newblock
\urldef\tempurl%
\url{https://doi.org/10.1145/2531602.2531659}
\showURL{%
\tempurl}


\bibitem[\protect\citeauthoryear{Yadid and Yahav}{Yadid and Yahav}{2016}]%
        {yadid2016extracting}
\bibfield{author}{\bibinfo{person}{Shir Yadid} {and} \bibinfo{person}{Eran
  Yahav}.} \bibinfo{year}{2016}\natexlab{}.
\newblock \showarticletitle{Extracting code from programming tutorial videos}.
  In \bibinfo{booktitle}{\emph{Proceedings of the 2016 ACM International
  Symposium on New Ideas, New Paradigms, and Reflections on Programming and
  Software}}. \bibinfo{publisher}{ACM}, \bibinfo{pages}{98--111}.
\newblock
\urldef\tempurl%
\url{https://doi.org/10.1145/2986012.2986021}
\showURL{%
\tempurl}


\bibitem[\protect\citeauthoryear{YouTube}{YouTube}{2020}]%
        {zoo}
\bibfield{author}{\bibinfo{person}{YouTube}.} \bibinfo{year}{2020}\natexlab{}.
\newblock \bibinfo{title}{Me at the zoo}.
\newblock
\newblock
\newblock
\shownote{Retrieved Sept 20, 2020 from \url{https://youtu.be/jNQXAC9IVRw}.}


\bibitem[\protect\citeauthoryear{Zhang and Cranshaw}{Zhang and
  Cranshaw}{2018}]%
        {zhang2018making}
\bibfield{author}{\bibinfo{person}{Amy~X. Zhang} {and} \bibinfo{person}{Justin
  Cranshaw}.} \bibinfo{year}{2018}\natexlab{}.
\newblock \showarticletitle{Making Sense of Group Chat Through Collaborative
  Tagging and Summarization}.
\newblock \bibinfo{journal}{\emph{Proc. ACM Hum.-Comput. Interact.}}
  \bibinfo{volume}{2}, \bibinfo{number}{CSCW}, Article \bibinfo{articleno}{196}
  (\bibinfo{date}{Nov.} \bibinfo{year}{2018}), \bibinfo{numpages}{27}~pages.
\newblock
\showISSN{2573-0142}
\urldef\tempurl%
\url{https://doi.org/10.1145/3274465}
\showDOI{\tempurl}


\end{thebibliography}
